\documentclass[12pt]{article}
\usepackage[margin=1in]{geometry}
\usepackage{authblk}
\usepackage{listings}
\usepackage{xcolor}
\usepackage{pifont}
\usepackage{float}
\usepackage{amsmath}
\usepackage[normalem]{ulem}
\usepackage{hyperref}
\hypersetup{
    colorlinks=true,
    linkcolor={red!50!black},
    citecolor={blue!50!black},
    urlcolor={blue!80!black},
}

\definecolor{backcolour}{rgb}{1,1,1}
\definecolor{codedark}{rgb}{.2,.2,1}
\definecolor{codelight}{rgb}{.7,.7,.7}
\definecolor{codegreen}{rgb}{0,.6,0}
\definecolor{codepurple}{rgb}{.58,0,.82}
\definecolor{highlight}{rgb}{.98,.97,.10}

\usepackage[lighttt]{lmodern} 

\lstdefinestyle{mystyle}{
    language=Python,
    backgroundcolor=\color{backcolour},   
    commentstyle=\color{codegreen},
    keywordstyle=\color{magenta},
    numberstyle=\tiny\color{codedark},
    stringstyle=\color{codepurple},
    basicstyle=\ttfamily\footnotesize,
    columns=fullflexible,
    frame=lines,
    showstringspaces=false,
    tabsize=4, 
    title=\lstname,
    numbers=none, 
    captionpos=b, 
    caption={},
    breakatwhitespace=false,         
    breaklines=true,                                   
    keepspaces=true,                 
    numbersep=5pt,                  
    showspaces=false,                
    showtabs=false,                  
    escapeinside={(@}{@)},
    moredelim=[is][\bfseries]{***}{***},
    moredelim=[is][\itshape]{___}{___},
    moredelim=[is][\color{lightgray}]{€€€}{€€€},
    moredelim=[is][\color{lightgray}\sout]{&&&}{&&&},
}
\usepackage{tikz}
\newcommand*\circled[1]{\tikz[baseline=(char.base)]{\node[shape=circle,draw,inner sep=2pt] (char) {#1};}}

\newcommand*\expand[1]{by modifying Listing~\ref{#1}}

\title{Diffusion Models to Enhance the Resolution of Microscopy Images: A Tutorial}

\author[1]{Harshith Bachimanchi\thanks{harshith.bachimanchi@physics.gu.se}}
\author[1]{Giovanni Volpe\thanks{giovanni.volpe@physics.gu.se}}
\affil[1]{Department of Physics, University of Gothenburg, Sweden}

\begin{document}

\maketitle

\begin{abstract}
Diffusion models have emerged as a prominent technique in generative modeling with neural networks, making their mark in tasks like text-to-image translation and super-resolution. 
In this tutorial, we provide a comprehensive guide to build denoising diffusion probabilistic models (DDPMs) from scratch, with a specific focus on transforming low-resolution microscopy images into their corresponding high-resolution versions. 
We provide the theoretical background, mathematical derivations, and a detailed Python code implementation using PyTorch, along with techniques to enhance model performance.\end{abstract}

\section{Introduction}

Over a century ago, the microscopist Ernst Abbe devised an equation showing how the  resolution of an optical microscope is limited by the wavelength of the illumination light~\cite{Abbe73_BeitraegeZur}. This critical limitation, known as the \emph{Abbe's diffraction limit}, implies that it is not possible to resolve objects smaller than 200 nanometers using an optical microscope. For scale, the diameter of a DNA molecule is about 2.5 nanometers---approximately one hundred times smaller. 
Since then, the quest to overcome this limit and to develop techniques for high-resolution imaging of cellular and subcellular structures has led to significant advancements in biomedical research~\cite{Hell94_BreakingDiffraction,Hell03_FluorescenceNanoscopy,Betzig06_ImagingIntracellular,Rust06_SubdiffractionlimitImaging,Gustafsson00_SurpassingLateral,Gustafsson05_NonlinearStructuredillumination} paving the way for \emph{super-resolution microscopy}. 
The super-resolution techniques that have revolutionized the field include structured illumination microscopy (SIM)~\cite{Gustafsson00_SurpassingLateral,Gustafsson05_NonlinearStructuredillumination}, stimulated emission depletion (STED)~\cite{Hell94_BreakingDiffraction,Hell03_FluorescenceNanoscopy}, stochastic optical reconstruction microscopy (STORM)~\cite{Rust06_SubdiffractionlimitImaging}, and photoactivated localization microscopy (PALM)~\cite{Betzig06_ImagingIntracellular}. However, these techniques require complex and expensive instrumentation, limiting their widespread availability.

Within the rapid growth of artificial intelligence (AI) research over the last decade, deep learning~\cite{LeCun15_DeepLearning,Volpe25_DeepLearningCrash} has shown promising results to increase the resolution of low-resolution images---a research area known as single-image super-resolution (SISR)~\cite{Yu24_ReviewSingle,Volpe25_DeepLearningCrash}.
The working principle typically involves training a deep neural network on matched pairs of low-resolution and high-resolution images to learn the unknown complex transformation between them without the need for analytical models~\cite{Qiao24_ZeroshotLearning}.
The SISR research field emerged from the photorealistic enhancement of everyday images~\cite{dong2015image,lim2017enhanced,lai2017deep,ledig2016photo,zhang2018image,yang2019deep,cai2019toward,cai2019ntire,lugmayr2020ntire,parmar2018image,menon2020pulse,dahl2017pixel,chen2018fsrnet} and has also recently been applied to enhance the resolution of microscopy images~\cite{Wang19_DeepLearning,Qiao21_EvaluationDevelopmenta,Qiao23_RationalizedDeep,Chen23_SingleframeDeeplearning,Qiao24_ZeroshotLearning}.
Notably, deep generative models such as generative adversarial networks (GANs) have succeeded to enhance the resolution of complex biological structures imaged under various microscopy modalities~\cite{Qiao21_EvaluationDevelopmenta,Wang19_DeepLearning,Qiao24_ZeroshotLearning}.

In recent years, diffusion models have been at the forefront of a new wave of generative modelling~\cite{Ho20_DenoisingDiffusion,Rombach22_HighResolutionImage}, having been applied for text-to-image and image-to-image translation tasks such as inpainting and super-resolution~\cite{Dhariwal21_DiffusionModels,Saharia22_PaletteImagetoImage,Rombach22_HighResolutionImage,li2022srdiff}.
There is also a growing trend within the scientific community to apply diffusion models across various fields, including microscopy~\cite{Maggiora23_ConditionalVariational,DellaMaggiora24_SingleExposure,Cross-Zamirski23_ClassGuidedImagetoImage} and other domains~\cite{Furrutter24_QuantumCircuit,lu2024diffusion}.

In this tutorial, we provide a step-by-step guide to implement diffusion models from scratch and apply them to image-to-image translation tasks.
We cover the essential theory, the mathematical derivations, and the practical code implementation. 
Even though we use super-resolution microscopy as an example, the diffusion models we describe in this tutorial can be applied to a broad range of image-to-image transformations.

\section{Computational Environment}

We start by setting up the computational environment necessary to run this tutorial.
The tutorial can be executed either on your local computer or using cloud-based platforms like Google Colab. In either case, the use of a GPU is highly recommended in order to execute the code within some reasonable time.
For local execution, it is recommended to create a fresh Python virtual environment with Python version 3.10 or higher where to install the required packages: \lstinline{torch}, \lstinline{torchvision}, \lstinline{torchmetrics}, \lstinline{tifffile}, \lstinline{matplotlib}, \lstinline{seaborn}, \lstinline{numpy}, \lstinline{tqdm}, \lstinline{deeplay}.
This can be done using the following pip command:
\begin{lstlisting}
pip install torch torchvision torchmetrics tifffile matplotlib seaborn numpy tqdm deeplay
\end{lstlisting}
Among these, the \lstinline{tifffile} package is used for reading and writing of TIF files, which are common in imaging tasks, and the \lstinline{tqdm} package provides graphical, extensible progress bars to track for loops and data processing pipelines.
The \lstinline{deeplay} package is a specialized library that extends PyTorch with additional functionalities and provides essential tools for deep learning workflows.
If you are running this tutorial on Google Colab, all the necessary packages except \lstinline{deeplay} are pre-installed. To install \lstinline{deeplay} on Google Colab, simply run \lstinline{!pip install deeplay} in the first code cell.

\section{Dataset}

We use the microtubule structure images of the BioSR dataset~\cite{Qiao21_EvaluationDevelopmenta}, which contains registered pairs of low-resolution and high-resolution images of four different biological structures. In increasing order of biological complexity, these structures are: clathrin-coated pits, endoplasmic reticulums, microtubules, and F-actin filaments.
The images are acquired through structured illumination microscopy (SIM)~\cite{Gustafsson00_SurpassingLateral,Gustafsson05_NonlinearStructuredillumination}. 
While in this tutorial we focus on the microtubule structure images, the implementation is identical for all structures and can be applied to other super-resolution datasets as well.

As the raw SIM images provided in the BioSR dataset require additional preprocessing, we have uploaded a processed version to a custom GitHub repository for convenience~\cite{DeepTrackAI_BioSR}. 
This repository contains 41,040 image pairs for training and 150 image pairs for testing across different noise levels. All images measure 128-by-128 pixels.

To reduce the computational demands of this tutorial, we have downscaled the original 256-by-256-pixel high-resolution images to 128-by-128 pixels, and the original 128-by-128-pixel low-resolution images to 64-by-64 pixels.
We then upsampled the low-resolution images to 128-by-128 pixels using bicubic interpolation.
This approach helps maintain consistency in image size for the deep learning models.
While the goal of this tutorial is to provide a comprehensive understanding of the process rather than focus on achieving benchmark results, we recommend checking the GitHub repository of the BioSR dataset for benchmarking purposes~\cite{GitHub_BioSR}.

\subsection{Downloading the BioSR Dataset}

Let us begin by downloading the dataset from the GitHub repository, as shown in Listing~\ref{cd:download_dataset}.
\begin{lstlisting}[    
    label=cd:download_dataset,
    caption=Downloading the BioSR dataset for microtubules
]
import os

if not os.path.exists("biosr_dataset"): (@\circled{1}@)
    os.system("git clone https://github.com/DeepTrackAI/biosr_dataset") (@\circled{2}@)
\end{lstlisting}
This code checks whether the dataset directory already exists \circled{1} and, if not, downloads the dataset from the GitHub repository \circled{2}.
If you are implementing this tutorial on Google Colab, the download should take approximately 8 minutes.
Alternatively, you can directly download the dataset from the repository and store it in your local machine, if you are executing the code on your computer.

\subsection{Implementing a Custom Dataset Class}

In PyTorch, we need to use a dataset object to efficiently handle and preprocess data, such as images, text, or other types of inputs. The dataset abstraction provides a way to load and transform data on the fly, which is essential for tasks involving large datasets that cannot be loaded into memory all at once. PyTorch provides a \lstinline{Dataset} class as a template for creating custom datasets tailored to specific needs.

A class that inherits from the \lstinline{Dataset} class needs to implement two key methods:
\begin{itemize}
    \item \lstinline{__len__()}: This method returns the total number of samples in the dataset, which is crucial for determining the length of the dataset.
    \item \lstinline{__getitem__(index)}: This method retrieves a sample from the dataset at the specified index. It is responsible for loading the data, applying any necessary transformations, and returning the processed data in the desired format.
\end{itemize}
By using these methods, PyTorch can efficiently manage large datasets by loading data in small batches during training, which helps in handling high-resolution images or other data-intensive tasks.

Thus, we will now create a custom class to load the images, apply the required transformations, and return the input-output image pairs, as shown in Listing~\ref{cd:dataset_class}.
\begin{lstlisting}[
    label=cd:dataset_class,
    caption=Dataset class to manage the images
]
import torch
from tifffile import tifffile

class BioSRDataset(torch.utils.data.Dataset):
    """Dataset class for lading the BioSR dataset."""

    def __init__(self, low_res_dir, high_res_dir, transform=None): (@\circled{1}@)
        """Initialize dataset."""
        self.low_res_dir = low_res_dir
        self.high_res_dir = high_res_dir
        self.transform = transform

        self.filename_list = [filename for filename 
                              in os.listdir(self.low_res_dir) 
                              if filename.endswith(".tif")] (@\circled{2}@)

    def __len__(self):
        """Return the number of image pairs."""
        return len(self.filename_list) (@\circled{3}@)

    def __getitem__(self, index):
        """Get next low-resolution--high-resolution image pair."""
        low_res_path = os.path.join(self.low_res_dir, 
                                    self.filename_list[index])
        low_resolution_image = tifffile.imread(low_res_path) (@\circled{4}@)

        high_res_path = os.path.join(self.high_res_dir, 
                                     self.filename_list[index])
        high_resolution_image = tifffile.imread(high_res_path) (@\circled{5}@)

        if self.transform: (@\circled{6}@)
            low_resolution_image = self.transform(low_resolution_image)
            high_resolution_image = self.transform(high_resolution_image)

        return low_resolution_image, high_resolution_image
\end{lstlisting}
This code implements the custom dataset class to load the images in small batches without having to load the entire dataset into memory before training. It is initialized with the following input parameters \circled{1}: \lstinline{low_res_dir} sets the directory for the low resolution images; \lstinline{high_res_dir} sets the directory for the high resolution images; and \lstinline{transform} contains the transformations to be applied to the images. The initialization method concludes by saving a list of available TIF files in the low-resolution image folder \circled{2}. 
The length of the dataset is set to the length of the file list \circled{3}.

The \lstinline{__getitem__()} method loads the images corresponding to a certain index \circled{4}\circled{5}, applies the transformations \circled{6}, and finally returns a list with the low-resolution and high-resolution images.

\subsection{Defining the Augmentations, Transformations, and Normalization}

When working with deep learning models, especially in image processing tasks, it is crucial to apply various data preprocessing techniques to enhance the model's performance and generalization capacity. Augmentations and transformations can help create a more diverse dataset by introducing variations in the images, which can prevent the model from overfitting. Normalization ensures that the input data is on a consistent scale, which can improve the convergence of the model during training. These preprocessing steps are essential for preparing the dataset in a way that maximizes the model's ability to learn effectively from the data.

We will now define the necessary augmentations, transformations, and normalization to be applied to the dataset, as shown in Listing~\ref{cd:transforms}.
\begin{lstlisting}[
    label=cd:transforms,
    caption={Implementing the necessary augmentations, transforms, and normalization}
]
from torchvision import transforms

transform = transforms.Compose([
    transforms.ToTensor(), (@\circled{1}@)
    transforms.Normalize((0.5,), (0.5,)) (@\circled{2}@)
])
\end{lstlisting}
This code defines a simple transformation, which first converts an image to a PyTorch tensor \circled{1} and then normalizes it to have the pixel values in the range of -1 and 1 \circled{2}.
Since the dataset already contains images augmented through flipping and random rotations, we will not add any augmentations to avoid repeating the same images.

\subsection{Creating the Training and Testing Datasets}

We now divide the data into training and testing sets.
The training dataset is used to teach the model the relevant patterns and relationships within the data. The testing dataset, on the other hand, is reserved for evaluating the model's performance on unseen data, ensuring that the model generalizes well to new inputs. Properly creating and managing these datasets is essential for robust model development and accurate performance assessment.

We will now create the training and testing datasets, as shown in Listing~\ref{cd:datasets}.
\begin{lstlisting}[
    label=cd:datasets,
    caption=Creating the datasets
]
root = os.path.join("biosr_dataset", "BioSR", "Microtubules") (@\circled{1}@)

train_dataset = BioSRDataset(low_res_dir=os.path.join(root, "training_wf"),
                             high_res_dir=os.path.join(root, "training_gt"),
                             transform=transform)
                             
test_dataset = BioSRDataset(low_res_dir=os.path.join(root, "test_wf", "level_09"), (@\circled{2}@)
                            high_res_dir=os.path.join(root, "test_gt"), (@\circled{3}@)
                            transform=transform) 
\end{lstlisting}
This code starts by defining the parent folder that contains the train and test datasets \circled{1}.
The datasets are then defined by setting the appropriate file paths for the low-resolution and high-resolutions images, and applying the transforms defined in Listing~\ref{cd:transforms}.

Both datasets consist of low-resolution images of varying noise levels, generated by exciting the biological samples using high-power lasers at different intensities, all within the same field of view.
In the test dataset, low-resolution images with different noise levels are organized into separate folders to evaluate the model performance efficiently for each noise level.
The folders are named according to their noise levels, with \lstinline{"level_01"} corresponding to low excitation and high noise, and \lstinline{"level_09"} representing high excitation and low noise.
The noise level is set to \lstinline{"level_09"}, indicating that images have low noise \circled{2}.
Alternatively, you can set the folder to a noise level of your choice.
This structuring does not apply to the high-resolution images, which are stored in a single folder, as there is only one ground truth high-resolution image corresponding to the various noise levels of the low-resolution images \circled{3}.

\subsection{Visualizing Some Low- and High-Resolution Images}

\begin{figure}[H]
    \centering
    \includegraphics[width=\textwidth]{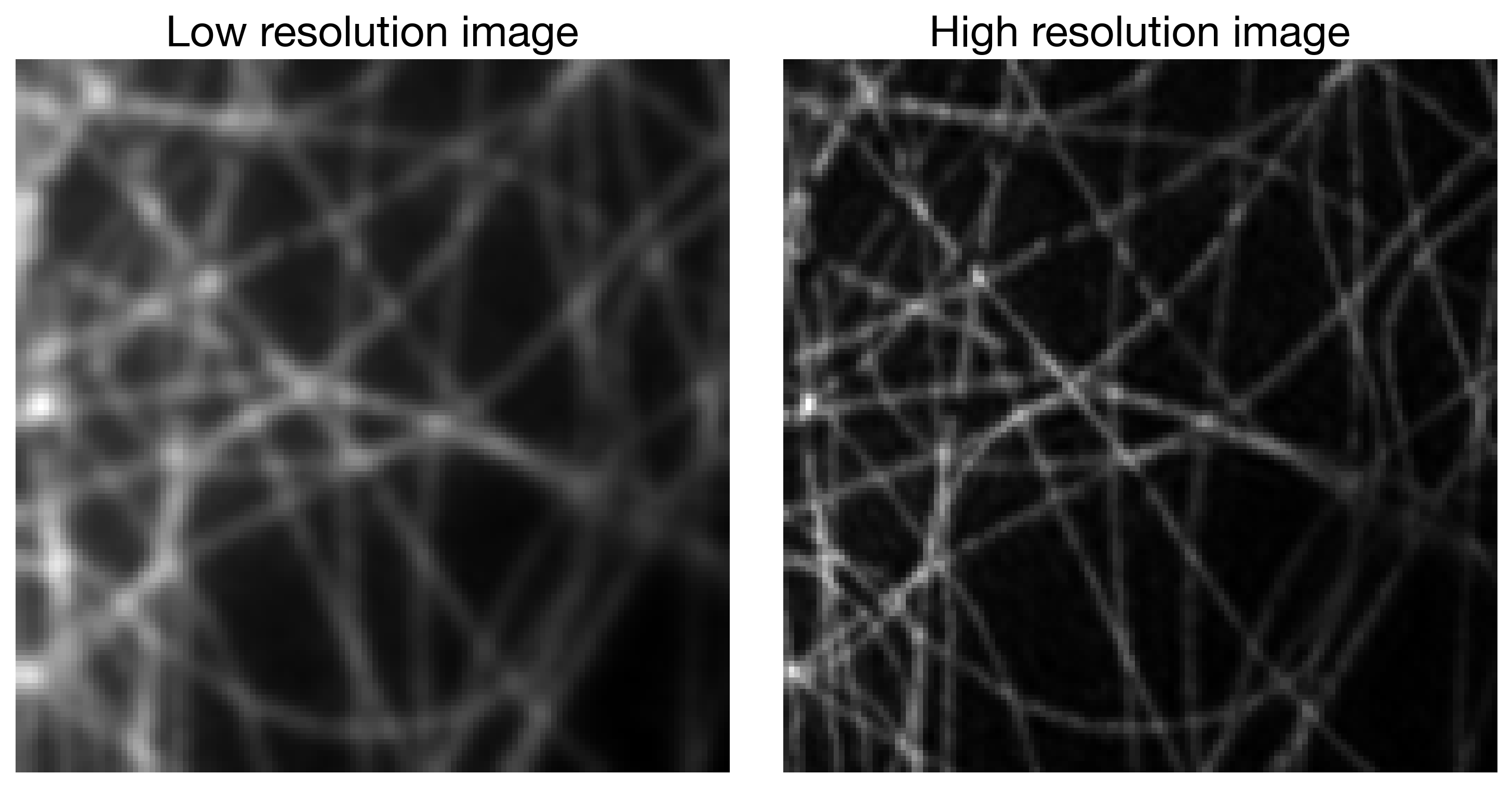}
    \caption{Low-resolution (left) and high-resolution (right) images of microtubules}
    \label{fig:sample_images}
\end{figure}

Let us now visualize some example images, using Listing~\ref{cd:plot_images}.
\begin{lstlisting}[
    label=cd:plot_images,
    caption=Plotting some low-resolution and high-resolution images
]
import matplotlib.pyplot as plt
import numpy as np

low_res_image, high_res_image = train_dataset[np.random.randint(0, len(train_dataset))] (@\circled{1}@)

plt.figure()

plt.subplot(1, 2, 1)
plt.imshow(low_res_image.permute(1, 2, 0), cmap="gray") (@\circled{2}@)
plt.title("Low-resolution image")
plt.axis("off")

plt.subplot(1, 2, 2)
plt.imshow(high_res_image.permute(1, 2, 0), cmap="gray") (@\circled{3}@)
plt.title("High-resolution image")
plt.axis("off")

plt.show()
\end{lstlisting}
This code samples a random low- and high-resolution image pair from the train dataset \circled{1}, and plots them side-by-side \circled{2}\circled{3}. An example of the resulting figure is shown in Fig.~\ref{fig:sample_images}.

As we can see, the low-resolution image is significantly blurred in comparison to its high-resolution counterpart, where the details of the microtubules are clearly visible.

\section{Diffusion Models for Super-Resolution}

Diffusion models are the latest torchbearers in the field of generative AI~\cite{Ho20_DenoisingDiffusion,Rombach22_HighResolutionImage,Volpe25_DeepLearningCrash}. These models, standing alongside other influential generative models like variational autoencoders (VAEs)~\cite{kingma2013auto,vahdat2020nvae} and generative adversarial networks (GANs)~\cite{goodfellow2014generative,karras2018progressive,radford2015unsupervised}, have quickly gained acclaim for their exceptional ability to generate high-quality images~\cite{Dhariwal21_DiffusionModels}. 
The scope of diffusion models extends far beyond mere image generation. These models have also excelled in tasks like unconditional/conditional image generation~\cite{Dhariwal21_DiffusionModels}, inpainting~\cite{Rombach22_HighResolutionImage}, and super-resolution~\cite{Saharia21_ImageSuperResolution}, often outperforming even the most advanced GANs. 

We will start by understanding the core principles of denoising diffusion probabilistic models (DDPMs), while simultaneously implementing the code with the super-resolution task in mind. 

\subsection{Denoising Diffusion Probabilistic Models (DDPM)}

DDPMs are generative models that aim to learn the probabilistic distribution underlying a given dataset in order to generate new and unseen images with the same statistical properties.
In contrast to VAEs, where the images/samples in the dataset are explicitly mapped to a pre-defined distribution in the latent space---the parameters of which are learned through training, diffusion models follow an implicit approach where the network understands the underlying data distribution by learning to denoise some synthetically added noise to the clean input images.

\begin{figure}[H]
    \centering
    \includegraphics[width=\textwidth]{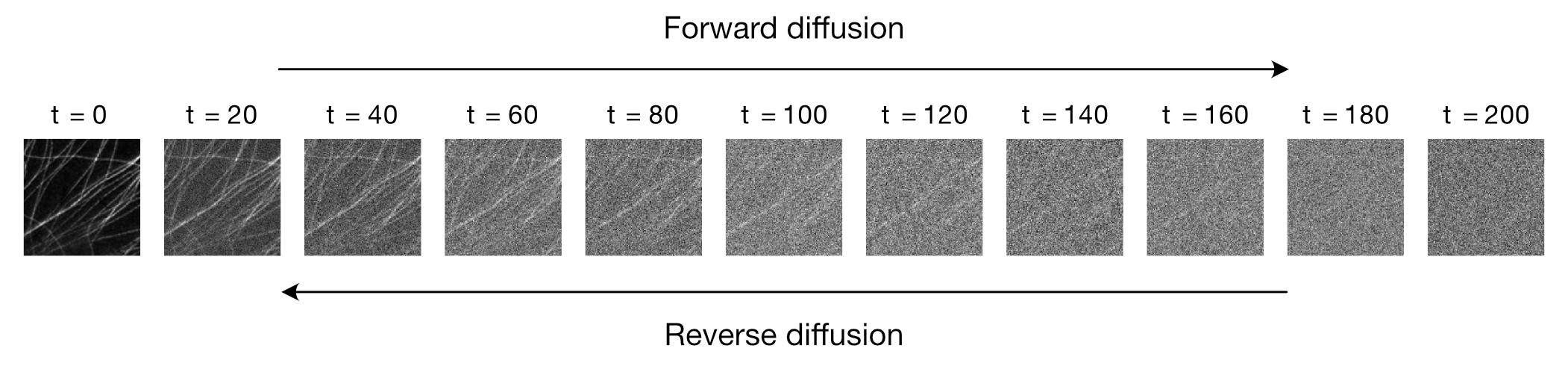}
    \caption{Illustration of forward and reverse diffusion processes for a high-resolution image from the BioSR dataset}
    \label{fig:diffusion}
\end{figure}

The \emph{forward diffusion process} involves the sequential addition of noise to an image until it is completely corrupted, as illustrated in Fig.~\ref{fig:diffusion} going from left to right. 
This is a Markov process in the space of all images, meaning that the next image is solely determined by the current one (and therefore it is independent from the previous ones).
In the forward process shown in Fig.~\ref{fig:diffusion}, the next image is just the current image plus some noise (and therefore does not depend explicitly on the previous images).

Conversely, the \emph{reverse diffusion process} goes in the opposite direction, successively removing noise from an image, as illustrated in Fig.~\ref{fig:diffusion} going from right to left. Clearly, the reverse diffusion is a much more challenging process, but, as we will see in this tutorial, we can train an artificial neural network to denoise the noisy image, gradually recovering a cleaner version of the image. 
In essence, by learning to denoise the images in the reverse process, the network implicitly learns about the images themselves.
We will see more on this in the next section, where we will derive the forward and reverse diffusion processes in their mathematical forms, and understand the connection to the neural network.

\subsection{The Forward Diffusion Process}\label{sec:forward_diffusion}

In the forward diffusion process, a small amount of Gaussian noise is added to an image iteratively until the image is completely corrupted with noise, as shown in Fig.~\ref{fig:diffusion} going from left to right.

Let us start with an image ${\bf x}_0$ sampled from a dataset $q({\bf x})$ that contains a collection of images.
The image ${\bf x}_0$ is a matrix filled with the values of the pixels.
The subscript $0$ denotes the time step $t = 0$, corresponding to the clean image from the dataset, which will be transformed into a completely noisy image as the time steps progress, $t = 0, 1, \dots, T$.

At each step in the forward diffusion process, we make the image slightly noisier by adding some Gaussian noise.
In probabilistic terms, the forward diffusion process can be written as
\begin{equation}\label{eq:dist_forward_diffusion}
    q({\bf x}_t | {\bf x_{t-1}})
    =
    \mathcal{N}( 
    \underbrace{\sqrt{1-\beta_t} \, {\bf x_{t-1}}}_{\mbox{$\boldsymbol{\mu}$}}, \, 
    \underbrace{\beta_t \, {\bf I}}_{\mbox{$\boldsymbol{\sigma}^2$}}
    )
\end{equation}
where $q({\bf x}_t | {\bf x_{t-1}})$ denotes the conditional probability of obtaining the image ${\bf x}_t$ given the previous image ${\bf x}_{t-1}$, represented as a Gaussian distribution $\mathcal{N}(\boldsymbol{\mu}, \boldsymbol{\sigma}^2)$ with mean $\boldsymbol{\mu} = \sqrt{1-\beta_t} \ {\bf x_{t-1}}$ dependent on the previous noisy image ${\bf x_{t-1}}$, and variance $\boldsymbol{\sigma}^2 = \beta_t {\bf I}$ with $\beta_t$ being a time-dependent parameter that controls the amount of noise added at each step (also known as the \emph{variance schedule}).

We can convert Eq.~\ref{eq:dist_forward_diffusion} into a recursive equation, using a useful property of Gaussian distributions known as \emph{reparametrization}.
This mathematical trick allows us to write any Gaussian distribution with mean $\boldsymbol{\mu}$ and variance $\boldsymbol{\sigma}^2$ in terms of the standard Normal distribution (i.e., a Gaussian distribution with mean ${\bf 0}$ and variance ${\bf 1}$). 
In more mathematical terms, a random variable ${\bf x} \sim \mathcal{N}(\boldsymbol{\mu}, \boldsymbol{\sigma}^2)$ can be written as ${\bf x} = \boldsymbol{\mu} + \boldsymbol{\sigma} \ \boldsymbol{\epsilon}$, where $\boldsymbol{\epsilon} \sim \mathcal{N}({\bf 0}, {\bf 1})$. 
Applying this property to the forward diffusion equation (Eq.~\ref{eq:dist_forward_diffusion}), we can write
\begin{equation}\label{eq:forward_diffusion}
    {\bf x}_{t+1} = \underbrace{\sqrt{1-\beta_t} \, {\bf x}_t}_{\mbox{rescaled image}} 
    +
    \underbrace{\sqrt{\beta_t} \, {\boldsymbol{\epsilon}}_t}_{\mbox{noise}}
\end{equation}
where ${\boldsymbol{\epsilon}}_t$ is a noise obtained from a Normal distribution (this is a matrix containing a random number for each pixel of the image). 
The subscript in ${\boldsymbol{\epsilon}}_t$ makes it clear that its value is different at each time step, even though it is always sampled from the same standard Normal distribution.
The higher the value of $\beta_t$, the more noise is added to the image at time step $t$. 

Using this equation, we can straightforwardly generate the noisy version of the image at time $t+1$ given the image at time $t$.
For example, for a clean image ${\bf x}_0$ from the BioSR dataset, we can generate its first noisy version ${\bf x}_1$, and iteratively use ${\bf x}_1$ to generate ${\bf x}_2$, then ${\bf x}_2$ to generate ${\bf x}_3$, and so on until ${\bf x}_T$.
This process can be repeated multiple times, so that from a single initial image, we can obtain multiple different images that ``diffuse'' in the vector space of all possible images.

\subsubsection{The Fast Forward Diffusion Process}

We can take advantage of the  Gaussian structure of the noise and the Markovian nature of the forward process to accelerate the computation of the noisy images. 
In fact, when training the neural network to model the reverse process, we need many examples of the forward process at different, randomly sampled time steps. 
Therefore, if $T$ is large as required for the DDPM models (often $T=1000$), this process quickly becomes cumbersome and computationally prohibitive, because we must generate all the intermediate noisy images to reach a noisy image at a specific time step.
Luckily, we can exploit the mathematical properties of the stochastic process described by Eq.~\ref{eq:forward_diffusion} to directly jump from the initial image to the image at time $t$, without having to generate all the intermediate noisy images:
\begin{equation}\label{eq:fastforward}
    {\bf x}_{t} = \sqrt{\bar\alpha_t} \, {\bf x}_0 + \sqrt{1-\bar\alpha_t} \, \boldsymbol{\epsilon}
\end{equation}
where $\bar\alpha_t = \prod_{s=1}^t \alpha_s$, $\alpha_t = 1 - \beta_t$, and $\boldsymbol{\epsilon} \sim \mathcal{N}({\bf 0}, {\bf 1})$.
The derivation of the fast forward process can be found in Appendix~\ref{appendix:fast_forward} and in references~\cite{Volpe25_DeepLearningCrash, weng2021diffusion}.

\subsection{Implementing the Forward Diffusion Process}\label{sec:implement_forward_diffusion}

We are now ready to implement the forward diffusion process in PyTorch to generate the noisy versions of the images. Before we do this, let us define the device on which the computations are performed, as shown in Listing~\ref{cd:device}. Setting the right device can significantly reduce the training time and improve performance.
\begin{lstlisting}[
    label=cd:device,
    caption=Determining the device to be used in the computations
]
if torch.cuda.is_available():
    device = torch.device("cuda:0") (@\circled{1}@)
elif torch.backends.mps.is_available():
    device = torch.device("mps") (@\circled{2}@)
else:
    device = torch.device("cpu") (@\circled{3}@)
\end{lstlisting}
This code selects a GPU if it is available \circled{1}\circled{2}, and otherwise it sets the device to CPU \circled{3}.
The statement \lstinline{"cuda:0"} selects the first GPU if multiple GPUs are available and is equivalent for most cases to just \lstinline{"cuda"} \circled{1}.
The statement \lstinline{"mps"} selects the Apple Silicon GPU device on MacOS, if available \circled{2}.
You can determine which device has been selected with the \lstinline{print(device)} command.

After having defined the device on which the computations are performed, we can now implement the forward diffusion process to generate noisy versions of images. We will do this by creating the custom \lstinline{Diffusion} class, as shown in Listing~\ref{cd:forward_diffusion}.
\begin{lstlisting}[
    label=cd:forward_diffusion,
    caption=Implementing the forward diffusion process
]
class Diffusion:
    """Denoising diffusion probabilstic model (DDPM)."""
    
    def __init__(self, noise_steps=1000, beta_start=1e-4, beta_end=0.02,
                 device=device):
        """Initialize the diffusion model."""
        self.noise_steps = noise_steps
        self.beta_start = beta_start
        self.beta_end = beta_end
        self.device = device

        self.beta = self.prepare_noise_schedule().to(device) (@\circled{1}@)
        self.alpha = 1.0 - self.beta (@\circled{2}@)
        self.alpha_bar = torch.cumprod(self.alpha, dim=0) (@\circled{3}@)

    def prepare_noise_schedule(self): (@\circled{4}@)
        """Prepare the noise schedule."""
        return torch.linspace(self.beta_start, self.beta_end, self.noise_steps)

    def forward_diffusion(self, x0, t): (@\circled{5}@)
        """Implement the forward diffusion process."""
        sqrt_alpha_bar = torch.sqrt(self.alpha_bar[t])[:, None, None, None]
        sqrt_one_minus_alpha_bar = \
            torch.sqrt(1 - self.alpha_bar[t])[:, None, None, None]
        noise = torch.randn_like(x0)

        return sqrt_alpha_bar * x0 + sqrt_one_minus_alpha_bar * noise, noise (@\circled{6}@)
\end{lstlisting}
This class uses Eq.~\ref{eq:fastforward} to implement the forward diffusion process. It takes the number of noise steps as input parameter, and prepares a noise schedule to modify clean input images with appropriate noise levels.

The \lstinline{__init__()} method takes the following input parameters: \lstinline{noise_steps} sets the number of noise steps; \lstinline{beta_start} and \lstinline{beta_end} set the beginning and the ending noise levels for the noise scheduler; and \lstinline{device} sets the device on which the operations are performed.

The noise schedule, \lstinline{beta} \circled{1}, is created with the help of the \lstinline{prepare_noise_schedule()} method \circled{4}, which takes the \lstinline{beta_start}, \lstinline{beta_end}, and \lstinline{noise_steps} parameters to generate a linear noise schedule.
It is then used to create the \lstinline{alpha} \circled{2} and \lstinline{alpha_bar} \circled{3} values, which are necessary for the fast forward process, as shown in Eq.~\ref{eq:fastforward}. 

Finally, the \lstinline{forward_diffusion()} method \circled{5} takes the input image \lstinline{x0} and the time step \lstinline{t} to return the noisy version of the image and noise itself by implementing the fast forward process \circled{6}.

Now, we will define an example instance of the \lstinline{Diffusion} class to generate some noisy images from a clean high-resolution image. This is shown in Listing~\ref{cd:forward_diffusion_example}.
\begin{lstlisting}[
    label=cd:forward_diffusion_example,
    caption=Sampling images in the forward diffusion process
]
diffusion = Diffusion(noise_steps=201, beta_start=0.0001, beta_end=0.02) (@\circled{1}@)

low_res_image, high_res_image = train_dataset[
    torch.randint(0, len(train_dataset), (1,)).squeeze()
] (@\circled{2}@)

time_steps = torch.arange(0, 201, 20).int() (@\circled{3}@)

noisy_images = []
for i in range(len(time_steps)):
    noisy_image, noise = diffusion.forward_diffusion( (@\circled{4}@)
        x0=high_res_image[None, ...].to(device), 
        t=torch.tensor([time_steps[i]]).to(device)
    )
    noisy_images.append(noisy_image)
\end{lstlisting}
This code starts by instantiating the diffusion class with $200$ noise steps, and with beta values ranging from $0.0001$ to $0.02$ \circled{1}. 
Then, it randomly samples a low- and high-resolution image pair from the training dataset \circled{2}.
It generates ten noisy images at intervals of $20$ time steps \circled{3} with the help of the \lstinline{forward_diffusion()} method, which takes the high-resolution image and the time step as inputs, and returns the noisy image \circled{4}. 
The noisy images are stored in the \lstinline{noisy_images} list, which we will use for plotting.

We can now visualize the noisy images and thereby the forward diffusion process with Listing~\ref{cd:visualize_forward_diffusion}.
\begin{lstlisting}[
    label=cd:visualize_forward_diffusion,
    caption=Visualizing the noisy images generated in the forward diffusion process
]
fig, axs = plt.subplots(1, len(time_steps))
for i, ax in enumerate(axs.flatten()):
    ax.imshow(noisy_images[i].cpu().numpy().squeeze(), cmap="gray")
    ax.set_title(f"t = {time_steps[i]}", fontsize=10)
    ax.axis("off")
plt.tight_layout()
plt.show()
\end{lstlisting}
This code extracts the noisy images generated in Listing~\ref{cd:forward_diffusion_example}, and plots them in a sequence. This results in the series of increasingly noisy images shown in Fig.~\ref{fig:diffusion}.

\subsection{The Reverse Diffusion Process}\label{sec:reverse_diffusion}

As we have already seen in the previous sections, the reverse diffusion process is the opposite of the forward diffusion process. 
The idea is to slightly reduce the amount of noise in an image in each time step. 
Therefore, over the whole process, an initial image that is just noise becomes a sharp, clean image, as in Fig.~\ref{fig:diffusion} going from right to left.
Let us see how we can approach this mathematically.

Since the forward process $q({\bf x}_t | {\bf x}_{t-1})$ (Eq.~\ref{eq:forward_diffusion}) is a Gaussian process, we can reasonably assume that the reverse process will have a similar structure, with the conditional probability reversed as $q({\bf x}_{t-1} | {\bf x}_t)$, which denotes the conditional probability of obtaining a slightly denoised image ${\bf x}_{t-1}$ given the current noisier image ${\bf x}_{t}$.
However, it turns out that this equation is intractable as it would require an understanding of the whole dataset, which we do not have.
More intuitively, since the goal of this process is to eventually obtain a clean image ${\bf x}_{0}$ that has the properties that characterize the images in the dataset, it is impossible to derive the specifics of the reverse process to make it tractable without having a prior knowledge about the dataset and its complex relationship with the noise.

However, the equation becomes tractable when it is conditioned on a clean image  ${\bf x}_0$, as now it has some new information about the dataset. This conditional distribution with additional new information, mathematically known as the \emph{posterior} $q({\bf x}_{t-1} | {\bf x}_{t}, {\bf x}_0)$ can be expressed as
\begin{equation}\label{eq:posterior_reverse_diffusion}
    q({\bf x}_{t-1} | {\bf x}_t, {\bf x}_0)
    =
    \mathcal{N}(
    \underbrace{\tilde{\boldsymbol{\mu}} ({\bf x}_t, {\bf x}_0)}_{\mbox{posterior mean}}, \
    \underbrace{\tilde{\beta_t} {\bf I}}_{\mbox{posterior variance}}
    )
\end{equation}
where
\begin{equation}\label{eq:posterior_mean}
    \tilde{\boldsymbol{\mu}} ({\bf x}_t, {\bf x}_0)
    =
    \frac{\sqrt{\bar{\alpha}_{t - 1}} \ \beta_t}{1 - \bar{\alpha}_t} \ {\bf x}_0
    +
    \frac{\sqrt{\alpha_t} (1 - \bar{\alpha}_{t-1})}{1 - \bar{\alpha}_t} \ {\bf x}_t
\end{equation}
and
\begin{equation}\label{eq:posterior_variance}
    \tilde{\beta_t}
    =
    \frac{1 - \bar{\alpha}_{t-1}}{1 - \bar{\alpha}_t} \ \beta_t
\end{equation}
Here, $\tilde{\boldsymbol{\mu}} ({\bf x}_t, {\bf x}_0)$, and $\tilde{\beta_t}$ are the posterior mean and variance.
Their derivation requires advanced probability concepts, which are beyond the scope of this tutorial, but can be found in the references~\cite{Ho20_DenoisingDiffusion, weng2021diffusion}.

We can simplify the posterior mean in Eq.~\ref{eq:posterior_mean} by rearranging the forward diffusion in Eq.~\ref{eq:forward_diffusion} to obtain
\begin{equation}\label{eq:rearrange_fastforward}
    {\bf x}_0 
    = 
    \frac{{\bf x}_{t} - \sqrt{1-\bar\alpha_t} \, {\boldsymbol{\epsilon}}}{\sqrt{\bar\alpha_t}}
\end{equation}
We can then plug ${\bf x}_0$ into Eq.~\ref{eq:posterior_mean} to derive a simplified form of posterior mean:
\begin{equation}\label{eq:simplified_posterior_mean}
    \tilde{\boldsymbol{\mu}} ({\bf x}_t, t)
    =
    \frac{1}{\sqrt{\alpha_{t}}}
    \left(
        {\bf x}_t \,
        -
        \frac{1 - \alpha_{t}}{\sqrt{1 - \bar{\alpha_{t}}}} \,
        {\boldsymbol{\epsilon}}_t
    \right)
\end{equation}
The derivation of this result can be found in Appendix~\ref{appendix:posterior_mean}.

This is a very valuable result as now we can use the mathematical forms of posterior mean (Eq.~\ref{eq:simplified_posterior_mean}) and variance (Eq.~\ref{eq:posterior_variance}) to write a recursive equation for the reverse process using the reparametrization trick, similar to the approach we followed for the forward process (Eq.~\ref{eq:forward_diffusion}). 
However, note that this equation is derived from the posterior (Eq.~\ref{eq:posterior_reverse_diffusion}), which comes with the critical limitation of its dependence on the initial image ${\bf x_0}$.
Since the main objective of the reverse process is to generate a clean image from pure noise, using a clean image as a starting condition contradicts the fundamental intent of this process.
This is where we will employ a neural network.

\subsubsection{Why Using a Neural Network?}

As a brief recap, we have seen that the reverse process becomes tractable when it is modelled with the posterior $q({\bf x}_{t-1} | {\bf x}_t, {\bf x}_0)$, whose mean and variance now have derivable mathematical expressions when conditioned on the clean image ${\bf x}_0$.
This very dependence, however, makes the reverse process impractical, as our goal is to arrive at the clean image ${\bf x}_0$ by systematically denoising a pure noise image ${\bf x}_T$.
We can overcome this limitation by making a neural network learn about the dataset, rather than directly providing the specific details about the dataset in the form of a clean image.
We will now see how this strategy works.

We begin by formulating the reverse process as $p_{\theta}({\bf x}_{t-1} | {\bf x}_t)$, which does not explicitly depend on the clean image ${\bf x}_0$ but relies on ${\theta}$, the internal parameters of a neural network. This relationship can be expressed as
\begin{equation}\label{eq:reverse_diffusion_nn}
    p_{\theta}({\bf x}_{t-1} | {\bf x}_t)
    =
    \mathcal{N}(
    \boldsymbol{\mu}_{\theta} ({\bf x}_t, t ), \
    \boldsymbol{\Sigma}_{\theta} ({\bf x}_t, t )
    )
\end{equation}
where the mean $\boldsymbol{\mu}_{\theta} ({\bf x}_t, t )$ and variance $\boldsymbol{\Sigma}_{\theta} ({\bf x}_t, t )$ are now estimated by the neural network, which learns the dataset structure implicitly.

Now we can directly compare $p_{\theta}({\bf x}_{t-1} | {\bf x}_t)$ with the posterior $q({\bf x}_{t-1} | {\bf x}_{t}, {\bf x}_0)$, aiming to make the neural network effectively model the tractable posterior. By comparing the means of both distributions, we obtain
\begin{equation}
    \boldsymbol{\mu}_{\theta} ({\bf x}_t, t )
    =
    \frac{1}{\sqrt{\alpha_{t}}}
    \left(
        {\bf x}_t \,
        -
        \frac{1 - \alpha_{t}}{\sqrt{1 - \bar{\alpha_{t}}}} \,
        {\boldsymbol{\epsilon}_{\theta}({\bf x}_t, t)}
    \right)
\end{equation}
Since we already have the input noisy image ${\bf x}_t$, we can just make the neural network predict the noise $\boldsymbol{\epsilon}_{\theta}$ added to the image (notice the subscript $\theta$ indicating that $\boldsymbol{\epsilon}_{\theta}$ is predicted by the neural network).

As for the variance, the authors of the DDPM~\cite{Ho20_DenoisingDiffusion} chose to fix it to an untrained time-dependent constant $\boldsymbol{\Sigma}_\theta({\bf x}_t, t) = \sigma^2_t {\bf I}$ so that, contrary to the mean, the variance is not estimated by the neural network. The variance could be set either to the forward process variance $\sigma^2_t = \beta_t$, or the posterior variance $\sigma^2_t = \tilde\beta_t$. 
Although both choices give similar results, we will use the first option for simplicity.

Now that we have the mathematical expressions for mean  $\boldsymbol{\mu}_{\theta} ({\bf x}_t, t )$ and variance $\boldsymbol{\Sigma}_\theta({\bf x}_t, t)$ in the reverse process $p_{\theta}({\bf x}_{t-1} | {\bf x}_t)$, we can use the reparametrization trick again to write the reverse equation in a recursive equation format as
\begin{equation}\label{eq:reverse_diffusion}
    {\bf x}_{t-1}
    =
    \frac{1}{\sqrt{\alpha_{t}}}
    \left(
        {\bf x}_t \,
        -
        \frac{1 - \alpha_{t}}{\sqrt{1 - \bar{\alpha_{t}}}} \,
        {\boldsymbol{\epsilon}_{\theta}({\bf x}_t, t)}
    \right)
    +
    \sqrt{\beta_t} \, {\bf z}
\end{equation}
where ${\bf z}$ is a noise term sampled from a standard Normal distribution (${\bf z} \sim \mathcal{N}(0, 1)$).
At each time step, the neural network predicts the noise ${\boldsymbol{\epsilon}_{\theta}({\bf x}_t, t)}$ present in the noisy image ${\bf x}_t$ given the time step $t$, and this process continues recursively from time steps $t=T$ to $0$.
Although ${\bf z}$ is also sampled from a Normal distribution like ${\bf x}_T$, it plays a different role.
While ${\bf x}_T$ represents the initial pure noise image that needs to be denoised, the term ${\bf z}$ is introduced at each step to account for the uncertainty in the model's predictions, ensuring that the reverse diffusion process remains flexible.
This flexibility prevents the process from being locked into a single deterministic path and allows it to explore a diverse range of possible clean images.
Additionally, the term is controlled by the $\sqrt{\beta_t}$ parameter, which effectively reduces the weight of ${\bf z}$ as the process progresses towards a clean image.

To train the neural network, the loss $\mathcal{L}$ is calculated by comparing the predicted noise ${\boldsymbol{\epsilon}_{\theta}({\bf x}_t, t)}$ with the noise $\boldsymbol{\epsilon}$ that is added to the image in the forward process through the mean squared error (MSE):
\begin{equation}
    \mathcal{L}
    =
    ||\boldsymbol{\epsilon} - \boldsymbol{\epsilon}_{\theta}({\bf x}_t, t)||^2
\end{equation}
To emphasize that the noisy image ${\bf x}_t$ in $\boldsymbol{\epsilon}_{\theta}({\bf x}_t, t)$ is originated from the forward diffusion process in Eq.~\ref{eq:fastforward}, we can further refine the loss function as
\begin{equation}
    \mathcal{L}
    =
    ||\boldsymbol{\epsilon} - \boldsymbol{\epsilon}_{\theta}(\sqrt{\bar\alpha_t} \, {\bf x}_0 + \sqrt{1-\bar\alpha_t} \, \boldsymbol{\epsilon}_t, t)||^2\\
\end{equation}
The selection of MSE as the loss function naturally arises when we consider how the diffusion models aim to match two probability distributions at each step: the model's predicted distribution $p_{\theta}({\bf x}_{t-1} | {\bf x}_t)$ and the true posterior distribution $q({\bf x}_{t-1} | {\bf x}_{t}, {\bf x}_0)$.
To bring these two distributions closer together, we minimize their difference using a measure called the Kullback-Leibler (KL) divergence. 
The KL divergence essentially quantifies how much one distribution diverges from the other. When applied to Gaussian distributions, minimizing the KL divergence simplifies to minimizing the MSE between the actual noise and the predicted noise at each step.
The full objective formulation behind the MSE loss involves some advanced probability concepts, such as variational lower bounds, and can be found in the references~\cite{Ho20_DenoisingDiffusion,weng2021diffusion}.
(The mean absolute error (MAE) also works well in many cases. Ultimately, the choice between MSE and MAE depends on the specific characteristics of the dataset and normalization.)

Before implementing of reverse process in code, let us take a moment to intuitively understand how the network learns about the data through denoising.
Remember that all that the network takes as input is the noisy image and the time step corresponding to the noise level of the image.
Additionally, the network learns information about the noise used to generate the noisy image, derived from the loss function.
During training, every time the network sees the same image from the dataset, it sees a different noisy version of that image, influenced both by the randomly sampled time step and by the noise added at that step.
As the network encounters the same image with varying levels of noise, and with the objective of predicting the noise in that image, over time it effectively learns the forward diffusion trajectory of the image for any randomly sampled noise.
This means that, given any randomly sampled noise, the network learns how to denoise it, as it now understands how an image of the dataset progresses through the noise space.

In essence, the network implicitly learns the data distribution by looking at the data through the lens of noise, much like how GANs learn through a competitive process.
However, this approach differs from other generative models such as the VAEs, where the data is mapped to a condensed latent space.
What truly sets the diffusion models apart is their ability to implicitly create an expansive and varied latent space embedded in a higher dimensional manifold.
By learning to navigate variations introduced by different sampled noises, the model encompasses an extensive spectrum of noise patterns and data variations.

\subsection{Implementing the Reverse Diffusion Process}\label{sec:implement_reverse_diffusion}

We are now ready to implement the reverse diffusion process, as described in Eq.~\ref{eq:reverse_diffusion}.
To do this, we will add a new method to the \lstinline{Diffusion} class, as shown in Listing~\ref{cd:reverse_diffusion}.
\begin{lstlisting}[
    label=cd:reverse_diffusion,
    caption=Implementing the reverse diffusion process (by expanding Listing~\ref{cd:forward_diffusion})
]
from tqdm import tqdm

€€€class Diffusion:
    """Denoising diffusion probabilstic model (DDPM)."""€€€
    
    €€€def __init__(self, noise_steps=1000, beta_start=1e-4, beta_end=0.02,€€€
        img_size=128,(@\circled{1}@) €€€device=device):
        . . . . . .€€€
        self.img_size = img_size

    €€€. . . . . .€€€

    def reverse_diffusion(self, model, n_images, n_channels, (@\circled{2}@)
                          position_encoding_dim, position_encoding_function,
                          save_time_steps=None, input_image=None):
        """Reverse diffusion process"""
        with torch.no_grad():
            x = torch.randn((n_images, n_channels, self.img_size, self.img_size)) (@\circled{3}@)
            x = x.to(self.device)
            
            denoised_images = []
            for i in tqdm(reversed(range(0, self.noise_steps)), (@\circled{4}@)
                          desc="U-Net inference", total=self.noise_steps):
                t = (torch.ones(n_images) * i).long() (@\circled{5}@)
                t_pos_enc = position_encoding_function( (@\circled{6}@)
                    t.unsqueeze(1), position_encoding_dim
                ).to(self.device)

                predicted_noise = model( (@\circled{7}@)
                    torch.cat((input_image.to(self.device), x), dim=1),
                    t_pos_enc,
                )

                alpha = self.alpha[t][:, None, None, None]
                alpha_bar = self.alpha_bar[t][:, None, None, None]
                
                if i > 0:
                    noise = torch.randn_like(x)
                else:
                    noise = torch.zeros_like(x)
                    
                x = (1 / torch.sqrt(alpha) * (x - ((1 - alpha) (@\circled{8}@)
                     / torch.sqrt(1 - alpha_bar)) * predicted_noise) 
                     + torch.sqrt(1 - alpha) * noise)

                if i in save_time_steps: (@\circled{9}@)
                    denoised_images.append(x)

            denoised_images = torch.stack(denoised_images)
            denoised_images = denoised_images.swapaxes(0, 1)
            return denoised_images
\end{lstlisting}
The \lstinline{Diffusion} class is now initialized with the additional parameter \lstinline{img_size} \circled{1}, which is used to control the image size in the reverse diffusion process. The default value is set to $128$ pixels, aligned with the specifications of the prepared BioSR dataset.

The new method \lstinline{reverse_diffusion()} implements the reverse diffusion process \circled{2}. 
It starts by creating a pure noise image sampled from a Gaussian distribution \circled{3}, where the shape of the image is defined by some input parameters: \lstinline{n_images} specifies the number of images to be sampled; \lstinline{n_channels} sets the number of image channels; and \lstinline{img_size} determines the spatial dimensions of the image (assumed to be square).

The image is then iteratively denoised through a for-loop, which runs from the set number of noise steps (\lstinline{noise_steps}) to $0$ \circled{4}. The for-loop is wrapped with the \lstinline{tqdm} function from the tqdm package to track the reverse diffusion process with a progress bar.

Within each iteration, a time step value for that iteration is created (the same as the loop variable \lstinline{i}) and is repeated to sample \lstinline{n_images} \circled{5}.
Next, the time step information is then passed through a position encoding function (given by the  \lstinline{position_encoding_function} parameter) to map the time step value to an encoded vector of a specified dimension (given by the parameter \lstinline{position_encoding_dim}) \circled{6}---we will see the position encoding function in detail in Listing~\ref{cd:position_encoding_function} in the next section.
Then, the noisy image and the encoded time step are passed through the neural network model \circled{7}, which tries to predict the noise that is applied to the image. 
To accommodate the reverse diffusion process for the image-to-image translation task, the input low-resolution image is concatenated to the noisy image before passing it to the model.
We will see the architecture and functioning of the model in Listing~\ref{cd:Attention_U-Net} in section~\ref{sec:attention_unet}.
Finally, with the predicted noise at hand, the noisy image is then cleaned, with the help of the reverse diffusion process \circled{8}, as defined by Eq.~\ref{eq:reverse_diffusion}. Notice the similarity between the code and the mathematical forms of the equation. 

This process is repeated for all the time steps, after which the noisy image is fully denoised and a clean image is generated.
To monitor the noisy image at different steps in the denoising process, the intermediate denoised images are stored into an empty list at regular time intervals defined by the \lstinline{save_time_steps} parameter \circled{9}.
Finally, the method returns the denoised images, including the final clean image and the intermediate images produced in the diffusion process.

\subsubsection{Alternative Implementation of the Reverse Process}

Let us now explore an alternative method for implementing the reverse diffusion process.
In the previous implementation, we have used the recursive equation (Eq.~\ref{eq:reverse_diffusion}) based on the simplified form of the posterior mean (Eq.~\ref{eq:simplified_posterior_mean}, Section~\ref{sec:reverse_diffusion} and Appendix~\ref{appendix:posterior_mean}).
As an alternative, we can directly use the original posterior mean in its unsimplified form (Eq.~\ref{eq:posterior_mean}) in the reverse process. 

In this approach, we need to reconstruct the clean image ${\bf x}_0$ at every time step using the forward diffusion equation (Eq.~\ref{eq:forward_diffusion}) and the noise predicted by the neural network.
This two-step approach (detailed in Appendix~\ref{appendix:alt_reverse_diffusion}) works better particularly with color images where the channel-wise information needs to be preserved across the transformation.
Therefore, if you are adapting this tutorial for color images, this alternative implementation may prove to be highly beneficial.

\subsubsection{Conditional DDPMs for Image-to-image translation tasks}\label{sec:conditional_generation}

As you might have already noticed, in the previous sections \ref{sec:implement_forward_diffusion} and~\ref{sec:implement_reverse_diffusion} the forward and reverse diffusion processes are solely applied to the high-resolution images.
This means that the DDPM is designed to understand the distribution of high-resolution images, and therefore, to sample unseen high-resolution images from pure noise.
However, since we are working with an image-to-image translation task, where the goal is to transform the images from low- to high-resolution, we provide the low-resolution image as a \emph{condition} during the reverse process, by passing it as an additional channel along with the noise (Section~\ref{sec:implement_reverse_diffusion}).
This is a way of conditioning the DDPM.

In traditional DDPM applications, where images are generated from pure noise, the forward diffusion process is directly applied to the images.
In conditional image generation tasks, such as generating images from class labels or text prompts, these conditions are provided as additional inputs to neural network during the reverse diffusion process
(We will briefly explore the technical modifications required to implement conditional DDPMs for class labels and text prompts in Section~\ref{sec:conclusion}).
For super-resolution tasks, where both the inputs and targets are images, we follow a similar approach by applying the forward diffusion process to the target high-resolution images.
The input low-resolution image is then provided a condition during the reverse diffusion process.

\subsection{Defining the Position Encoding Function}

We will now implement the position encoding function that is used to convert the time steps from their single-value representation to vectors of specified dimensions. The vectorial representations help the neural network to grasp the time step information in a more refined way. We will use a sinusoidal position encoding function, as shown in Listing~\ref{cd:position_encoding_function}.
\begin{lstlisting}[
    label=cd:position_encoding_function,
    caption=Position encoding function
]
def positional_encoding(t, enc_dim):
    """Encode position information with a sinusoid."""
    inv_freq = 1.0 / (10000 ** (torch.arange(0, enc_dim, 2).float() (@\circled{1}@)
                                / enc_dim)).to(t.device)
    pos_enc_a = torch.sin(t.repeat(1, enc_dim // 2) * inv_freq) (@\circled{2}@)
    pos_enc_b = torch.cos(t.repeat(1, enc_dim // 2) * inv_freq) (@\circled{3}@)
    pos_enc = torch.cat([pos_enc_a, pos_enc_b], dim=-1) (@\circled{4}@)
    return pos_enc
\end{lstlisting}
This sinusoidal position encoding function takes the time step, \lstinline{t}, and the dimension of the encoded vector, \lstinline{enc_dim}, as inputs and returns a unique vector corresponding to the time step.

It starts by creating an array of inverse frequencies by selecting the even indices from $0$ to the encoding dimension, dividing each by the encoding dimension, and then exponentiating the result with a base of 10,000 \circled{1}. This process generates a sequence of decreasing frequencies for the sine and cosine functions.

Next, the function repeats the value of the time step to match half of the encoding dimension (\lstinline{enc_dim // 2}), and multiplies this repeated array with the inverse frequencies to prepare the inputs for both sine and cosine functions.
It then applies the sine function on these inputs to create the first half of the position encoding \circled{2}, and the cosine function to create the latter half \circled{3}. Finally, these two halves are concatenated along the last dimension to form a complete position encoding vector corresponding to the time step \circled{4}.

\subsubsection{Generating and Visualizing the Position Encodings}

To gain some intuition on the position encoding process, we will now generate the position encodings of different time steps using Listing~\ref{cd:generate_position_encdoings}.
\begin{lstlisting}[
    label=cd:generate_position_encdoings,
    caption=Generating sinusoidal position encodings for different time steps
]
position_encoding_dim = 256 (@\circled{1}@)

pos_encs = []
for i in range(0, 100):
    t = torch.tensor([i])
    pos_enc = positional_encoding(t, position_encoding_dim) (@\circled{2}@)
    pos_encs.append(pos_enc.squeeze()) (@\circled{3}@)
pos_encs = torch.stack(pos_encs) (@\circled{4}@)
\end{lstlisting}
This code starts by setting the dimension of the encoded vector to $256$ \circled{1}. Then, it generates the position encodings for time steps from 0 to 100 in a for-loop \circled{2}. The position encodings are stored in an empty list \circled{3} and then stacked into a torch tensor \circled{4}. 

We can now visualize the position encodings using Listing~\ref{cd:visualize_position_encdoings}.
\begin{lstlisting}[
    label=cd:visualize_position_encdoings,
    caption=Visualizing the position encodings
]
fig = plt.figure()
plt.imshow(pos_encs.cpu().numpy(), cmap="Blues)
plt.xlabel("Encoding dimension")
plt.ylabel("Time step (t)")
plt.title("Sinusoidal position encodings")
plt.show()
\end{lstlisting}

\begin{figure}[H]
    \centering
	\includegraphics[width=\textwidth]{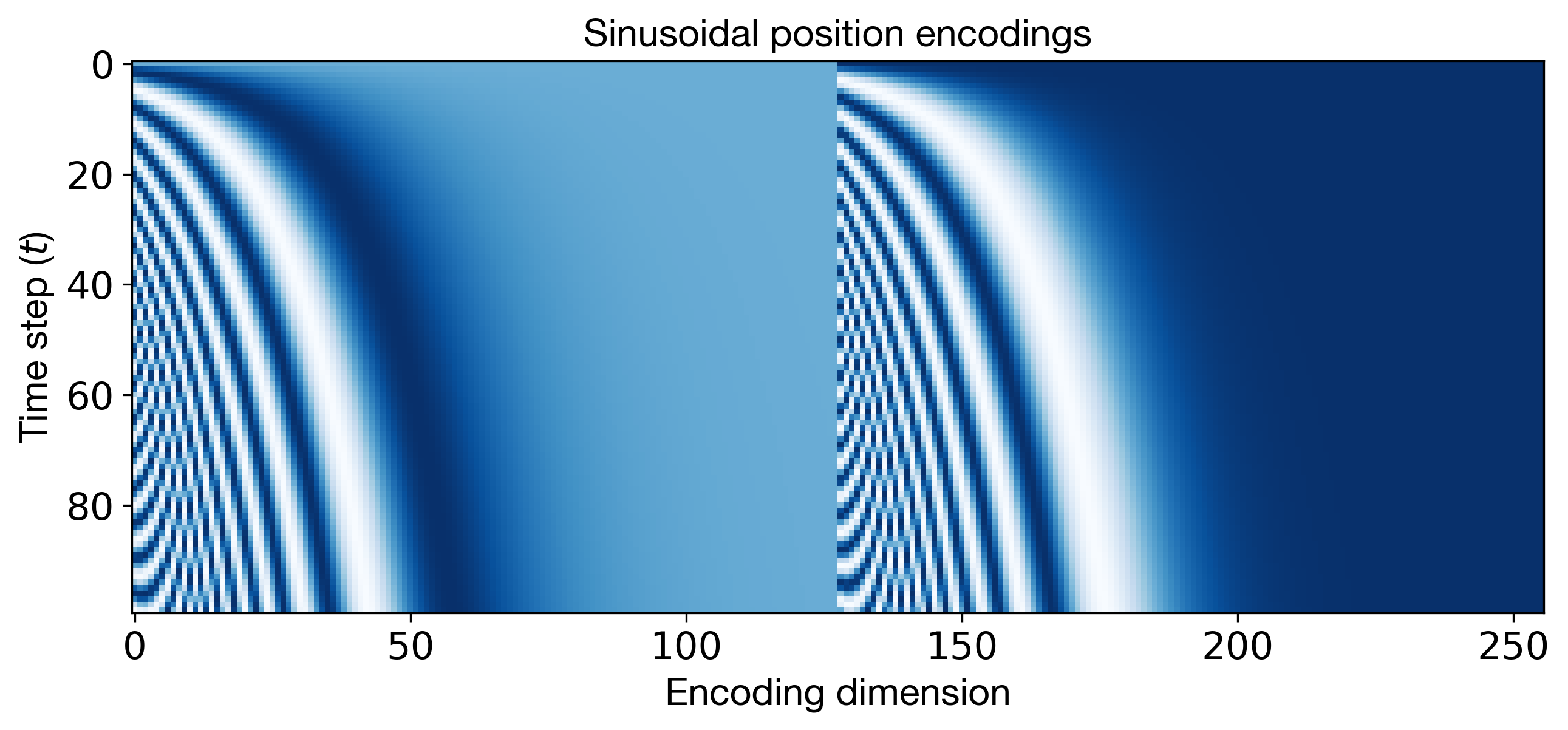}
	\caption{Sinusoidal position encodings for time steps 0 to 100}
	\label{fig:Position_encodings}
\end{figure}

The resulting image is shown in Fig.~\ref{fig:Position_encodings}. 

Each row in the image represents the position encoding of a specific time step.
Each encoding has a unique signature given by the alternating patterns of the sine and cosine functions within each vector, here represented on the left and right sides of the image respectively. Notice how the left and right sides of the image exhibit complementary patterns, together forming a unique encoding signature for each time step.

\subsection{Defining the Attention U-Net}\label{sec:attention_unet}

\begin{figure}[H]
    \centering
	\includegraphics[width=\textwidth]{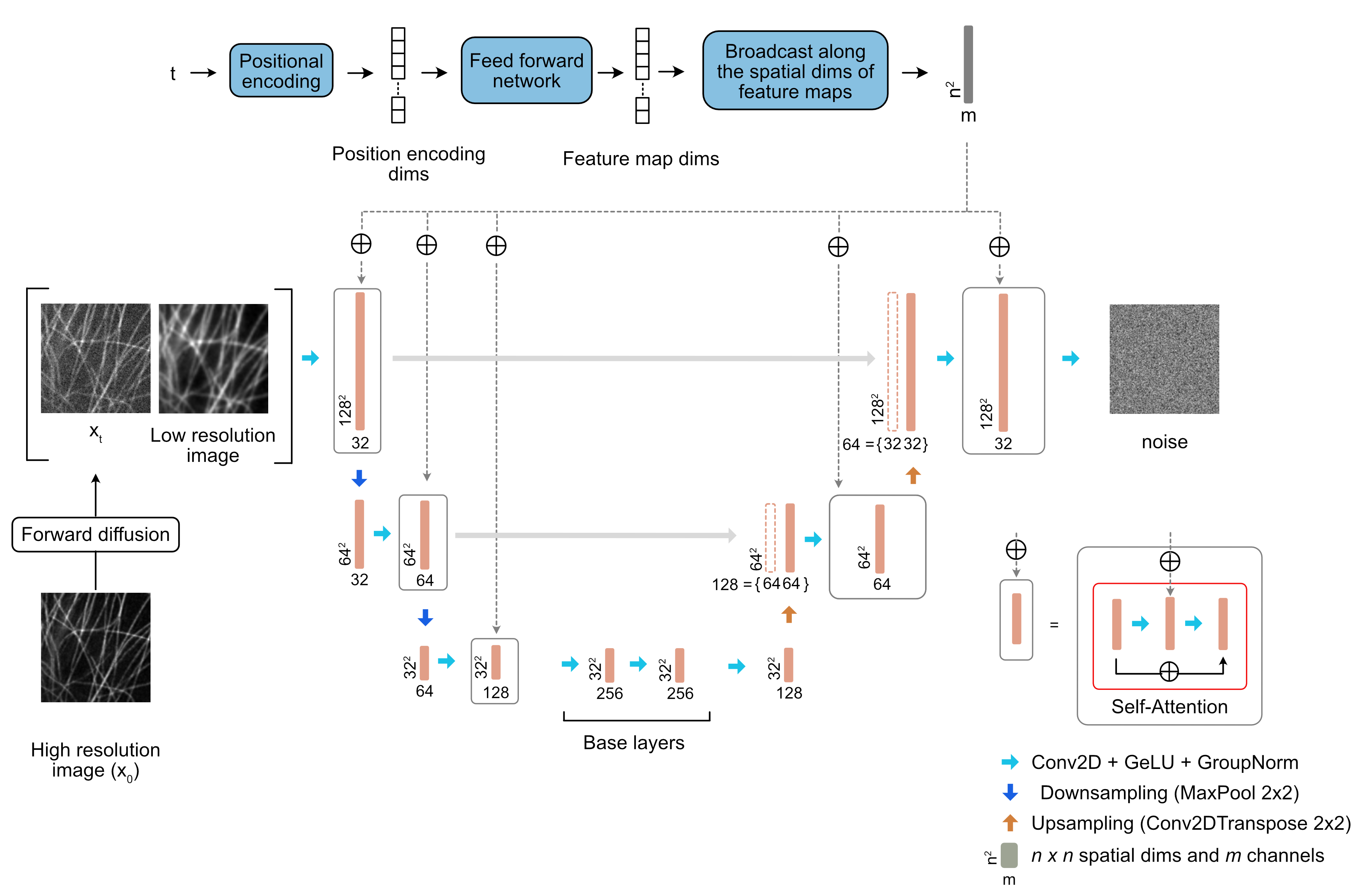}
	\caption{Attention U-Net architecture}
	\label{fig:Attention_U-Net}
\end{figure}

We will now define the architecture that we will use for the diffusion model. We will employ an attention U-Net, which is structurally similar to a traditional U-Net~\cite{ronneberger2015u} but with feature maps wrapped with attention layers to integrate the time step information~\cite{Vaswani17_AttentionAll}. An illustration of the attention U-Net architecture is shown in Fig.~\ref{fig:Attention_U-Net}.

The attention U-Net takes the noisy high-resolution image and the time step encoding (corresponding to the noisy image) as inputs, and outputs the noise applied to the image.
Here, the noisy image is obtained through the forward diffusion process applied on the high-resolution image (section~\ref{sec:forward_diffusion}).
Additionally, the noisy image is concatenated with the low-resolution image as a condition, to enable the image-to-image translation task.
The predicted noise is then processed by the reverse diffusion process, as described in the previous section.
Once the diffusion model is trained, the attention U-Net starts from a pure noise image concatenated with the low-resolution image and recursively denoises it in a loop to obtain a high-resolution image.

We will use the attention U-Net model from the Deeplay Python package (a deep learning library that extends PyTorch and Lightning with additional functionalities focused on code modularity and reusability~\cite{DeepTrackAI_deeplay}) as shown in Listing~\ref{cd:Attention_U-Net}.
\begin{lstlisting}[
    label=cd:Attention_U-Net,
    caption=Defining the attention U-Net model
]
import deeplay as dl

position_encoding_dim = 256

unet = dl.AttentionUNet( (@\circled{1}@)
    in_channels=2, (@\circled{2}@)
    channels=[32, 64, 128], (@\circled{3}@)
    base_channels=[256, 256], (@\circled{4}@)
    channel_attention=[False, False, False], (@\circled{5}@)
    out_channels=1, (@\circled{6}@)
    position_embedding_dim=position_encoding_dim, (@\circled{7}@)
)
unet.build() (@\circled{8}@)
unet.to(device) (@\circled{9}@)
\end{lstlisting}
This code instantiates the \lstinline{AttentionUNet} \circled{1}, builds it \circled{8}, and finally moves it to the device \circled{9}.

The number of input channels is set to 2 \circled{2} to account for the two inputs, namely the noisy version of the high-resolution image and the conditional low-resolution image. 
The output channel is set to 1 \circled{6} corresponding to the noise. 
The model is then initiated with three convolutional layers with 32, 64, and 128 filters \circled{3}. This also automatically sets the filters in the upsampling layers to the same numbers.
At the bottleneck, the U-Net is initialized with two additional convolutional layers with 256 and 256 filters \circled{4}.
Note that the number of filters in downsampling, upsampling, and bottleneck layers can be reduced to diminish the computational load, but at the cost of some loss in the performance of the network.

Unlike the traditional U-Net, the attention U-Net in diffusion models takes also the time step encoding as input. The time step encodings are fed into the model at different layers as shown in Fig.~\ref{fig:Attention_U-Net}. 
The encodings are reshaped to match the dimensions of the feature maps through dense layers and then added to the feature maps through residual connections. To facilitate these operations, the model also takes the dimension of the time step encoding vector as input, which in this case is set to 256 \circled{7}. 

A self-attention mechanism is then applied to the resulting feature map to help the model better interpret the relationship between the noisy image and the corresponding time step. It is however important to note that the attention mechanism can be computationally heavy. Therefore, the \lstinline{channel_attention} parameter can be used to control the attention mechanism applied at different levels \circled{5}. The boolean values (\lstinline{True} or \lstinline{False}) switch the attention on or off at the channels defined in the \lstinline{channels} parameter. 
Here, all the channel attentions are set to \lstinline{False} \circled{2} to minimize computational memory usage, especially since the images are large.
If sufficient memory is available, enabling one or more attentions can enhance the model's ability to better integrate the relationship between the time step and the noise image, thereby improving the model performance.

You can see the details of the attention U-Net architecture using \lstinline{print(unet)}.

\subsection{Training the Diffusion Model}

We are now ready to train the diffusion model for the super-resolution task. 
First, we will define the data loaders, loss function, and optimizer.
Then, we will instantiate the diffusion class and implement the training loop.

\subsubsection{Defining the Data Loaders}

We will define two different data loaders for the train dataset and for the test dataset, as shown in Listing~\ref{cd:data_loaders}.
\begin{lstlisting}[
    label=cd:data_loaders,
    caption=Defining the data loader
]
from torch.utils.data import DataLoader

batch_size = 64
train_loader = DataLoader(train_dataset, batch_size=batch_size, shuffle=True)
test_loader = DataLoader(test_dataset, batch_size=batch_size, shuffle=False)
\end{lstlisting}
This code defines train and test data loaders using a batch size of 64, and shuffles the data before creating batches for the training data loader. You can adjust the batch size depending on the computational resources at hand, with larger batch sizes improving performance and smaller batch sizes reducing the computational load.

\subsubsection{Defining the loss function}

We will now define the loss function, as shown in Listing~\ref{cd:loss_function}.
\begin{lstlisting}[
    label=cd:loss_function,
    caption=Defining the loss function
]
criterion = torch.nn.MSELoss()
\end{lstlisting}
This code defines a mean squared error loss (MSE) which will be applied between the noise predicted by the model and the actual noise added to the image.

\subsubsection{Defining the Optimizer}

We will now define the optimizer as shown in Listing~\ref{cd:optimizer}.
\begin{lstlisting}[
    label=cd:optimizer,
    caption=Defining the optimizer
]
optimizer = torch.optim.AdamW(unet.parameters(), lr=1e-4)
\end{lstlisting}
This code defines the Adam optimizer for the model with a learning rate set to 0.0001 through the parameter \lstinline{lr}.

\subsubsection{Instantiating the Diffusion Class}

Now, we will instantiate the \lstinline{Diffusion} class with the parameters required for the training, as shown in Listing~\ref{cd:instantiate_diffusion}.
\begin{lstlisting}[
    label=cd:instantiate_diffusion,
    caption=Instantiating the diffusion class
]
noise_steps = 2000

diffusion = Diffusion(
    noise_steps=noise_steps, (@\circled{1}@)
    img_size=128, (@\circled{2}@)
    beta_start=1e-6, (@\circled{3}@)
    beta_end=0.01, (@\circled{4}@)
)   
\end{lstlisting}
This code initiates a diffusion class with 2,000 noise steps \circled{1}, with image size set to 128 \circled{2} and with noise values ranging from 0.000001 \circled{3} to 0.01 \circled{4}.

\subsubsection{Implementing a Helper Function to Prepare Data}

Before implementing the training loop, we will create a helper function, \lstinline{prepare_data()}, which takes the input (low-resolution image) and the target (high-resolution image) from the dataset, and returns three outputs:
an image that contains the low-resolution image in the first channel and the noisy version of the high-resolution image at a randomly sampled time step in the second channel; 
the time step position encoding; and finally 
the added noise. 
Together these constitute the input-output pairs for training the model. 
The implementation of this function is shown in Listing~\ref{cd:prepare_data}.
\begin{lstlisting}[
    label=cd:prepare_data,
    caption=Function to prepare the data for super-resolution task
]
def prepare_data(input_image, target_image, device=device):
    """Prepare data."""
    batch_size = input_image.shape[0]

    input_image = input_image.to(device) 
    target_image = target_image.to(device) 

    t = torch.randint(low=0, high=noise_steps, size=(batch_size,)).to(device) (@\circled{1}@)

    x_t, noise = diffusion.forward_diffusion(target_image, t) (@\circled{2}@)
    x_t = torch.cat((input_image, x_t), dim=1) (@\circled{3}@)

    t = positional_encoding(t.unsqueeze(1), position_encoding_dim) (@\circled{4}@)

    return x_t.to(device), t.to(device), noise.to(device) (@\circled{5}@)
\end{lstlisting}
This function starts by creating a batch of random time step values that fall between 0 and the total number of noise steps \circled{1}. The target high-resolution image is then passed through the \lstinline{forward_diffusion()} method, which returns the noisy version of the high-resolution image and the added noise \circled{2}.
The input low-resolution image is concatenated to the noisy version of the high-resolution image \circled{3}. The time step is then sent to the position encoding function to get the time step encoding \circled{4}. Finally, all the outputs are sent to the device and returned as outputs of the function \circled{5}.

\subsubsection{Implementing the Training Loop}

We will now implement the training loop to teach the diffusion model to generate high-resolution from low-resolution images as shown in Listing~\ref{cd:training_loop}.
\begin{lstlisting}[
    label=cd:training_loop,
    caption=Training a conditional diffusion model for super-resolution task
]
epochs = 60 (@\circled{1}@)

for epoch in range(epochs):
    unet.train()

    for batch_idx, (input_images, target_images) in enumerate(train_loader, start=0):
        x_t, t, noise = prepare_data(input_images, target_images) (@\circled{2}@)

        outputs = unet(x=x_t, t=t) (@\circled{3}@)

        optimizer.zero_grad() (@\circled{4}@)
        loss = criterion(outputs, noise) (@\circled{5}@)
        loss.backward() (@\circled{6}@)
        optimizer.step() (@\circled{7}@)
\end{lstlisting}
This code trains the attention U-Net for the super-resolution task. The loop runs for 60 epochs \circled{1}, iterating over the batches of low- and high-resolution images that are obtained from the training data loader. For each batch, the images are passed through the \lstinline{prepare_data()} function \circled{2}, which outputs the noisy image (noisy high-resolution image concatenated with low-resolution image), the time step encoding, and the total noise added to the image up to the current time step.

The model then takes the noisy image and the time step encoding as inputs, and generates an output which represents the estimated noise present in the image \circled{3}.

The training process begins by zeroing the gradient buffers of the optimizer \circled{4}.
Next, the loss is calculated between the predicted noise and the actual noise \circled{5}. 
This loss is then backpropagated through the attention U-Net, which calculates the gradients with respect to each internal parameter of the network \circled{6}. 
Finally, the optimizer updates the parameters in a direction that minimizes the loss \circled{7}.

\subsection{Adding Timing and Notifications During the Training}

Now, we will add some quality-of-life improvements to the training loop to enhance the training experience and to get some real-time feedback by including timing and notifications within the training loop to understand the model's performance. Listing~\ref{cd:timing} implements this functionality modifying Listing~\ref{cd:training_loop}.
\begin{lstlisting}[
    label=cd:timing,
    caption=Timing the training and notifying the user (\expand{cd:training_loop}),
]
import time
from datetime import timedelta

€€€. . . . . .€€€

train_loss = [] (@\circled{1}@)

€€€for epoch in range(epochs):€€€
    start_time = time.time() (@\circled{2}@)
    num_batches = len(train_loader)

    print("\n" + f"Epoch {epoch + 1}/{epochs}" + "\n" + "_" * 10)

    €€€. . . . . .€€€

    running_loss = 0.0 (@\circled{3}@)
    €€€for batch_idx, (input_image, target_image) in enumerate(train_loader, start=0):€€€

        €€€. . . . . .€€€

        if batch_idx % 200 == 0:
            print(f"Batch {batch_idx + 1}/{num_batches}: " (@\circled{4}@)
                  + f"Train loss: {loss.item():.4f}")
        running_loss += loss.item() (@\circled{5}@)

    train_loss.append(running_loss / num_batches)(@\circled{6}@)
    end_time = time.time() (@\circled{7}@)

    print("-" * 10 + "\n" + f"Epoch {epoch + 1}/{epochs} : " (@\circled{8}@)
          + f"Train loss: {train_loss[-1]:.4f}, " 
          + f"Time taken: {timedelta(seconds=end_time - start_time)}")
\end{lstlisting}
This code records the time before \circled{2} and after each epoch \circled{7}, calculating the time elapsed between the two.
To keep track of the loss, it initializes an empty list to store the average training loss in every epoch \circled{1}. The average loss is computed by updating the running loss \circled{3} for every batch \circled{5} and dividing it by the total number of batches \circled{6}.
Throughout the code, print statements are added to keep track of the loss as the training progresses \circled{4}\circled{8}.

\subsubsection{Plotting the Intermediate Results During Training}

To gain further insights into the training process, we can now visualize the high-resolution images generated by the model for a given set of low-resolution images at the end of each epoch.
Listing~\ref{cd:intermediate_evaluation} adds this functionality to Listing~\ref{cd:timing}.
\begin{lstlisting}[
    label=cd:intermediate_evaluation,
    caption=Plotting sample high-resolution images after each epoch (\expand{cd:timing})
]
€€€. . . . . .€€€
€€€for epoch in range(epochs):€€€
    €€€. . . . . .€€€
    unet.eval() (@\circled{1}@)
    n_images = 4 (@\circled{2}@)
    
    for batch_idx, (test_input_images, (@\circled{3}@)
                    test_target_images) in enumerate(test_loader):
        generated_images = diffusion.reverse_diffusion ((@\circled{4}@)
            model=unet, (@\circled{5}@)
            n_images=n_images, (@\circled{6}@)
            n_channels=1, (@\circled{7}@)
            position_encoding_function=positional_encoding, (@\circled{8}@)
            position_encoding_dim=position_encoding_dim, (@\circled{9}@)
            input_image=test_input_images[:n_images], (@\circled{10}@)
            save_time_steps=[0], (@\circled{11}@)
        )
        break (@\circled{12}@)
    
    fig = plt.figure(figsize=(5, 8))
    
    for i in range(n_images):
        image_diff_trajectory = generated_images[i]

        low_res_image = test_input_images[i]
        generated_high_res_image = image_diff_trajectory[-1]
        target_high_res_image = test_target_images[i]

        plt.subplot(n_images, 3, i * 3 + 1)
        plt.title("Input")
        plt.imshow(low_res_image.permute(1, 2, 0).cpu().numpy(), cmap="gray") (@\circled{13}@)
        plt.axis("off")
        
        plt.subplot(n_images, 3, i * 3 + 2)
        plt.title("Output")
        plt.imshow(generated_high_res_image.permute(1, 2, 0).cpu().numpy(),
                   cmap="gray") (@\circled{14}@)
        plt.axis("off")
        
        plt.subplot(n_images, 3, i * 3 + 3)
        plt.title("Target")
        plt.imshow(target_high_res_image.permute(1, 2, 0).cpu().numpy(), 
                   cmap="gray") (@\circled{15}@)
        plt.axis("off")

    plt.show()
    plt.close(fig)
\end{lstlisting}
At the end of every epoch, this code sets the model to evaluation mode \circled{1}. 
This ensures that the model's behavior is consistent during evaluation by disabling certain layers and operations, such as dropout and batch normalization, which behave differently during training and testing phases.
For example, during training, batch normalization calculates statistics per batch and updates running statistics, while dropout randomly deactivates neurons.
During evaluation, when the model is set to evaluation model, batch normalization uses the fixed statistics calculated during training instead of recalculating them on test images, and drop out is turned off.
Although the attention U-Net does not use these layers, if you choose to modify the layers of the model's architecture to include them, it is essential to set the model to evaluation mode to maintain consistency.

It then draws a single batch of low-resolution and high-resolution images from the test dataset by iterating over the test loader \circled{3}.
Since we are in the training phase and just want to quickly evaluate the model, the iteration is interrupted after one batch with the \lstinline{break} command \circled{12}.

It then generates a given number of high-resolution images \circled{2}\circled{6} from low-resolution images \circled{10} through the reverse diffusion process implemented in the \lstinline{reverse_diffusion()} method \circled{3}. Recall that the reverse diffusion is applied on a concatenated image that contains pure noise and the low-resolution image.
This reverse diffusion method takes as input parameters the model \circled{5}, the number of images \circled{6}, the number of channels \circled{7}, the position encoding function \circled{8}, the encoding dim \circled{9}, the low-resolution images \circled{10}, and the the time steps at which the images are to be saved from the reverse diffusion process \circled{11}. 
This latter input is in the \lstinline{save_time_steps} parameter, which we set to a list containing only the last time step in the reverse diffusion process, 0, which corresponds to a fully denoised high-resolution image.
The tensors of the generated images will have the shape \lstinline{[n_images, save_time_steps, 1, 128, 128]}, where each image can be accessed via the first dimension, and the diffusion trajectory for each image can be accessed via the second dimension.
In this case, since we are only sampling the final clean image \circled{11}, the second dimension will contain only one image.

The code then plots the input low-resolution image \circled{13}, the generated high-resolution image \circled{14}, and the target high-resolution image \circled{15} side by side for visualization and comparison.

\begin{figure}[H]
    \centering
    \includegraphics[width=.7\textwidth]{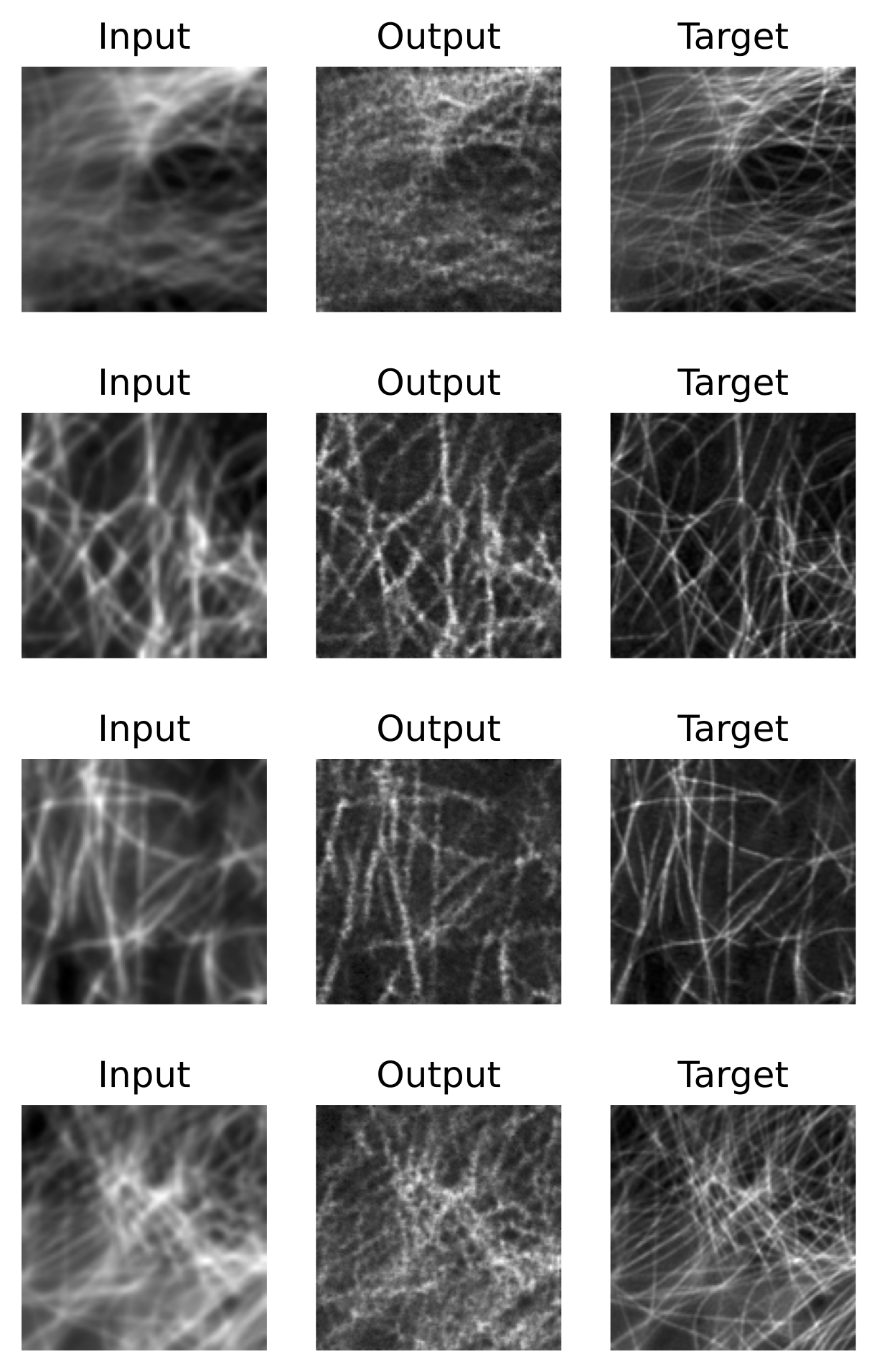}
    \caption{Input low-resolution images (left), generated high-resolution images after one training epoch (center), and target high-resolution images (right)}
    \label{fig:epoch_1}
\end{figure}

Fig.~\ref{fig:epoch_1} shows the output after a single epoch of training.

After one training epoch, the generated high-resolution images (center in Fig.~\ref{fig:epoch_1}) are still unstructured, containing a fair amount of noise when compared to the target high-resolution images (right). However, they are visually already better than the low-resolution images (left), which is a good sign indicating the training is progressing as expected.

\begin{figure}[H]
    \centering
    \includegraphics[width=.7\textwidth]{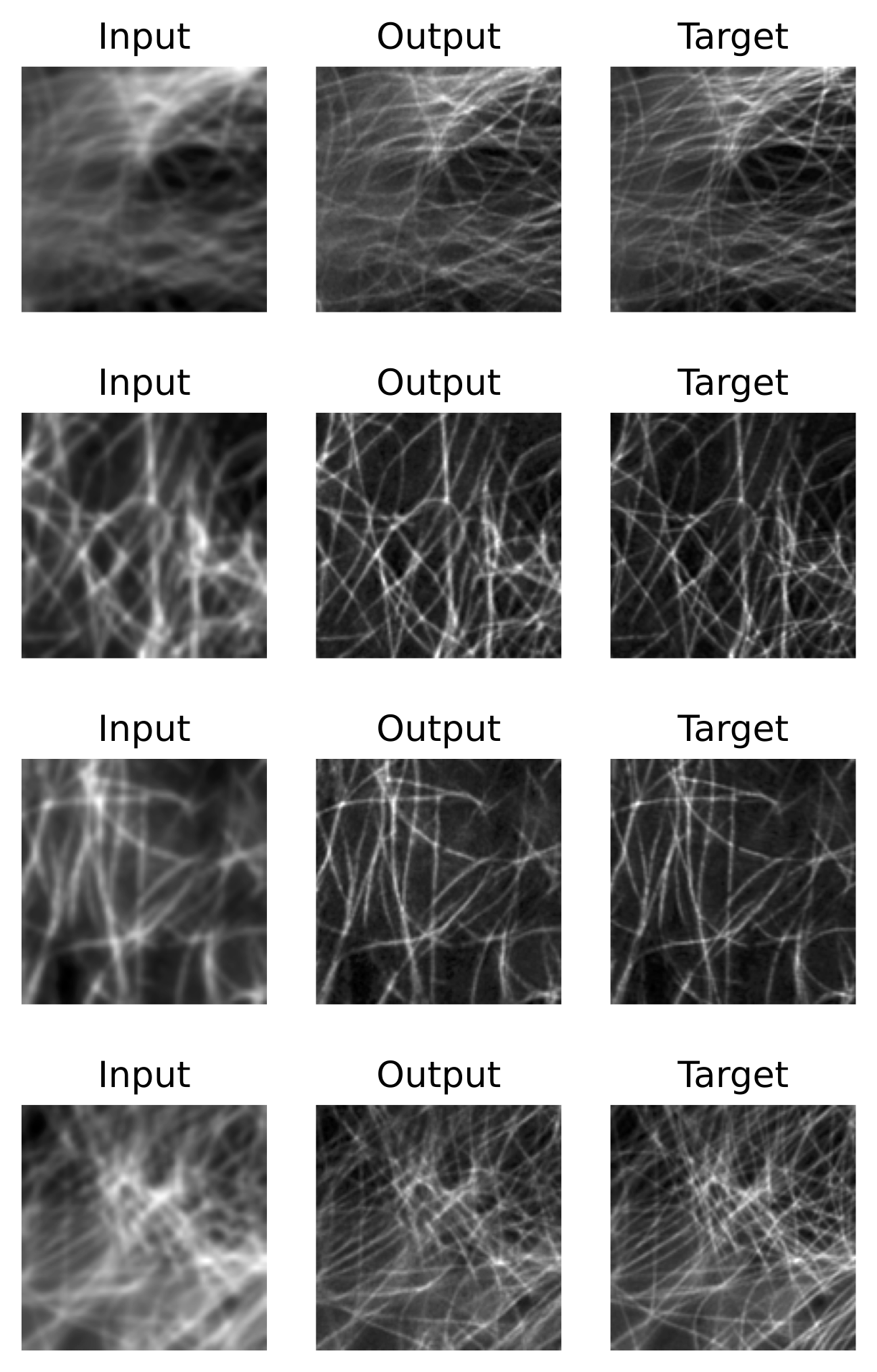}
    \caption{Input low-resolution images (left), generated high-resolution images after 60 training epochs (center), and target high-resolution images (right)}
    \label{fig:epoch_60}
\end{figure}

The generated images after 60 epochs are shown in Fig.~\ref{fig:epoch_60}.

After 60 training epochs, the generated high-resolution images (center in Fig.~\ref{fig:epoch_60}) are now visually much closer to the target high-resolution images (right), demonstrating a clear improvement over the input low-resolution images (left).

\subsubsection{Plotting the Training Losses}

\begin{figure}[H]
    \centering
    \includegraphics[width=\textwidth]{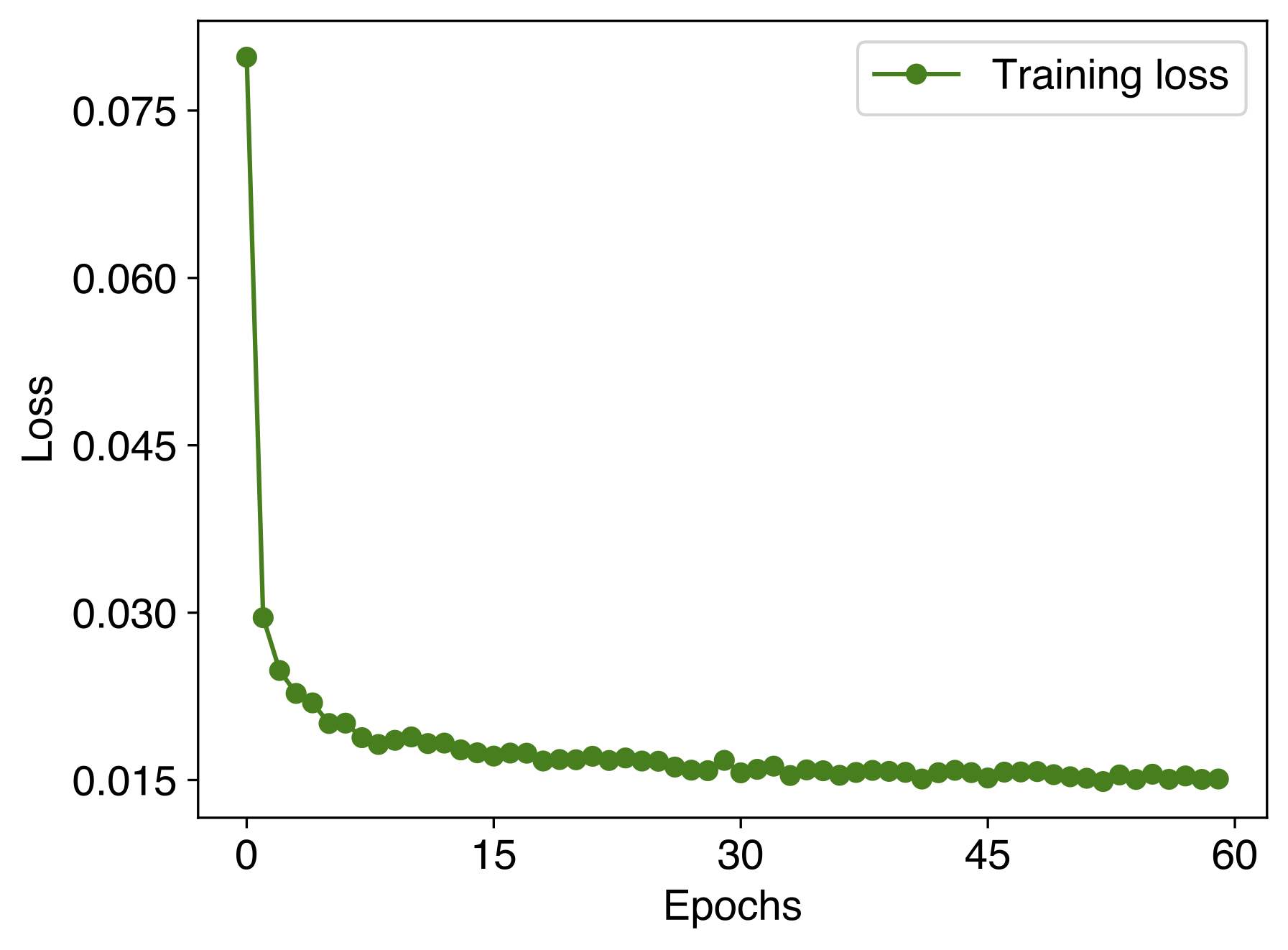}
    \caption{Evolution of training loss}
    \label{fig:train_loss}
\end{figure}

By analyzing the training loss, we can gain insights about the dynamics of the training process. 
Listing~\ref{cd:train_loss} illustrates how to plot the training loss to visualize the training behavior. 
\begin{lstlisting}[
    label=cd:train_loss,
    caption=Plotting the training loss
]
plt.figure()
plt.plot(train_loss, "g-o", label="Training loss")
plt.xlabel("Epochs")
plt.ylabel("Loss")
plt.legend()
plt.show()
\end{lstlisting}
This code generates Fig.~\ref{fig:train_loss}, which showcases the loss values over the course of training.

The loss is steadily decreasing over time, indicating that the training is progressing as expected. There is a downward trend also towards the end of the training indicating that the model is still learning. We can expect that, with further training, the loss can be further reduced, thus improving the quality of the generated images. 

Training with the above parameters takes approximately 5 minutes per epoch on a Nvidia T4 GPU in Google Colab compared to roughly 1 minute on an A100 GPU.
The training time can be further reduced by implementing several strategies: distributing the data across multiple GPU devices, which enhances parallel processing; increasing the batch size, which lowers the number of updates per epoch; reducing the complexity of the model, which decreases computations per step, and finally decreasing the number of steps in the diffusion process, which shortens the overall training duration and the inference time during generation.

\subsubsection{Evaluating the Model Performance}

We can now evaluate the model's performance more quantitatively by comparing the generated high-resolution images with the target images using various metrics. 
The metrics we will employ include the structural similarity index (SSIM), which assesses the perceptual quality of images by comparing luminance, contrast, and structure, and the multiscale structural similarity index (MS-SSIM), which extends SSIM by evaluating perceptual quality across multiple scales by downsampling the image and measuring SSIM at each scale. 
Both the metrics range from $-1$ to $1$, with $1$ indicating the highest perceptual similarity.

Additionally, we will consider pixel-to-pixel comparison metrics such as the the peak signal-to-noise ratio (PSNR), mean absolute error (MAE), and normalized root mean square error (NRMSE).
MAE and NRMSE are straightforward in their operation, directly measuring the differences between the generated and target images.
PSNR, on the other hand, measures the ratio of the maximum possible signal---the highest pixel value in the image---to the noise, which is the mean squared error between the generated and target images.
This ratio is expressed in decibels (dB), essentially quantifying the difference in terms of noise.
The higher the PSNR value, the better the quality of the generated image.
We will evaluate the performance of these metrics on the test dataset and monitor their evolution as the training progresses.
Listing~\ref{cd:metrics} adds this functionality to Listing~\ref{cd:intermediate_evaluation}.

\begin{lstlisting}[
    label=cd:metrics,
    caption=Evaluating the model performance (by modifying Listing~\ref{cd:intermediate_evaluation}) 
]
from torchmetrics.image import MultiScaleStructuralSimilarityIndexMeasure as MS_SSIM
from torchmetrics.image import StructuralSimilarityIndexMeasure as SSIM
from torchmetrics.image import PeakSignalNoiseRatio as PSNR
from torchmetrics.regression import MeanAbsoluteError as MAE

ms_ssim_metric = MS_SSIM( (@\circled{1}@)
    data_range=2.0, kernel_size=7, betas=(0.0448, 0.2856, 0.3001)
).to(device)
ssim_measure = SSIM(data_range=2.0).to(device) (@\circled{2}@)
psnr_measure = PSNR(data_range=2.0).to(device) (@\circled{3}@)
mae_measure = MAE().to(device) (@\circled{4}@)

€€€. . . . . .€€€

# Lists to store results (@\circled{5}@)
mae_results = []
ms_ssim_results = []
ssim_results = []
psnr_results = []
nrmse_results = []
€€€for epoch in range(epochs):€€€
    €€€. . . . . .€€€
    €€€unet.eval()€€€

    # Running losses to keep track of metrics in every batch (@\circled{6}@)
    running_mae = 0.0
    running_ms_ssim = 0.0
    running_ssim = 0.0
    running_psnr = 0.0
    running_nrmse = 0.0
    
    €€€for batch_idx, (test_input_images, \
                    test_target_images) in enumerate(test_loader):€€€
                    
        num_batches = len(test_loader)
        test_input_images = test_input_images.to(device)
        test_target_images = test_target_images.to(device)
        €€€. . . . . .€€€
        
        generated_images = diffusion.reverse_diffusion(
            €€€model=unet,€€€
            n_images=test_input_images.shape[0], (@\circled{7}@)
            €€€n_channels=1,
            position_encoding_dim=position_encoding_dim,
            position_encoding_function=positional_encoding,€€€
            input_image=test_input_images, (@\circled{8}@)
            €€€save_time_steps=[0],
        )€€€

        &&&break&&& (@\circled{9}@)

        generated_images_reshaped = generated_images.swapaxes(0, 1)[0] (@\circled{10}@)

        # Calculating the metrics for each batch (@\circled{11}@)
        batch_mae = mae_metric(generated_images_reshaped, test_target_images)
        batch_ms_ssim = ms_ssim_metric(generated_images_reshaped, test_target_images)
        batch_ssim = ssim_metric(generated_images_reshaped, test_target_images)
        batch_psnr = psnr_metric(generated_images_reshaped, test_target_images)
        batch_nrmse = torch.sqrt(
            torch.mean((generated_images_reshaped - test_target_images) ** 2)
        ) / (test_target_images.max() - test_target_images.min())

        # Accumulating the metrics (@\circled{12}@)
        running_mae += batch_mae.item()
        running_ms_ssim += batch_ms_ssim.item()
        running_ssim += batch_ssim.item()
        running_psnr += batch_psnr.item()
        running_nrmse += batch_nrmse.item()

    # Storing the mean metric values per epoch to the empty lists (@\circled{13}@)
    mae_results.append(running_mae / num_batches)
    ms_ssim_results.append(running_ms_ssim / num_batches)
    ssim_results.append(running_ssim / num_batches)
    psnr_results.append(running_psnr / num_batches)
    nrmse_results.append(running_nrmse / num_batches)
    €€€. . . . . .€€€
\end{lstlisting}
This code first imports the metrics from \lstinline{torchmetrics} and then defines instances of these metrics.
It instantiates the MS-SSIM metric \circled{1} with a data range of 2.0, which corresponds to the difference between the maximum and minimum values in the image, specifically $1$ and $-1$.
The kernel size is set to $7$, and the betas are set to standard values of $0.0448$, $0.2856$, $0.3001$.
Since MS-SSIM calculates the SSIM at different resolutions by downsampling the image, the betas represent the weights applied to the SSIM scores at each of three levels.
It then instantiates the SSIM \circled{2} and PSNR \circled{3} metrics with the data range parameter, and finally, it instantiates the MAE metric \circled{4}. The NRMSE metric is calculated directly through its mathematical formula in the code at the end.

The remainder of the code follows a similar flow as Listing~\ref{cd:intermediate_evaluation}.
In each epoch, it iterates over the test loader to generate high-resolution images from the low-resolution images.
The \lstinline{n_images} parameter is set to the number of images in the batch \circled{7}, and the \lstinline{input_image} parameter receives all the low-resolution images in the batch \circled{8}.
The \lstinline{break} command is omitted \circled{9}, ensuring that the process continues beyond a single batch and completes the entire test dataset.

The generated images are then reshaped for batch processing \circled{10}, after which the images are compared with the target images from the test dataset using the metrics instantiated earlier \circled{11}.
After each batch, the running metrics  \circled{6} are updated with the new values \circled{12}, and the average of all batches is then appended  \circled{13} to the corresponding lists defined for each metric \circled{5}.

Finally, at the end of training, the lists are populated with the metric values from each epoch, which can then be used for visualization.

Note that evaluating metrics during each epoch adds extra computational overhead. To speed up the training process, you can comment out the metric evaluation and run it separately on the test dataset once the model is fully trained.

\subsubsection{Plotting the Metrics}

We can now plot the metrics. By analyzing their evolution over the training process, we can assess the model's performance on the test dataset, which can help identify issues such as overfitting. Listing~\ref{cd:plot_metrics} demonstrates how to plot the metrics to visualize their behaviour on the test dataset.

\begin{figure}[H]
    \centering
    \includegraphics[width=\textwidth]{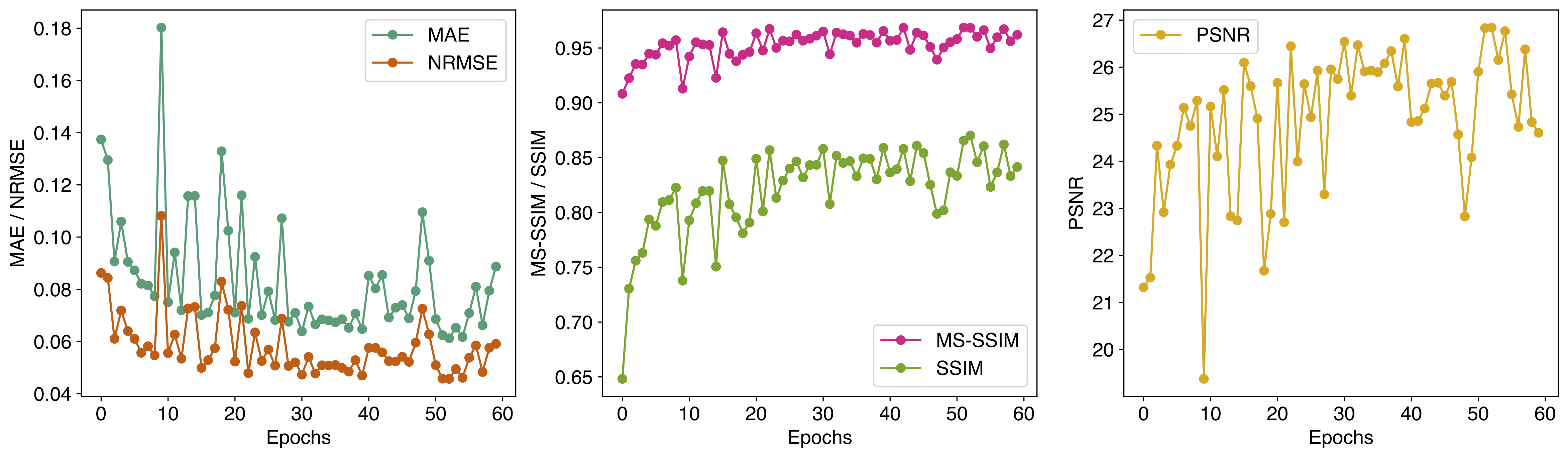}
    \caption{Evolution of metrics}
    \label{fig:metrics_evolution}
\end{figure}

\begin{lstlisting}[
    label=cd:plot_metrics,
    caption=Plotting the metrics
]
import seaborn as sns

palette = sns.color_palette("Dark2")

fig, ax = plt.subplots(1, 3, figsize=(19,5))

ax[0].plot(mae_results, color=palette[0], marker="o", label="MAE")        
ax[0].plot(nrmse_results, color=palette[1], marker="o", label="NRMSE")   
ax[0].set_xlabel("Epochs")
ax[0].set_ylabel("MAE / NRMSE")
ax[0].legend()

ax[1].plot(ms_ssim_results, color=palette[3], marker="o", label="MS-SSIM")  
ax[1].plot(ssim_results, color=palette[4], marker="o", label="SSIM")        
ax[1].set_xlabel("Epochs")
ax[1].set_ylabel("MS-SSIM / SSIM")
ax[1].legend()

ax[2].plot(psnr_results, color=palette[5], marker="o", label="PSNR")
ax[2].set_xlabel("Epochs")
ax[2].set_ylabel("PSNR")
ax[2].legend()

plt.show()
\end{lstlisting}
This code generates Fig.~\ref{fig:metrics_evolution} showcasing the evolution of metrics.

The MAE and NRMSE are steadily decreasing over time, indicating that the model is not overfitting to the training dataset and is generalizing well to the test dataset.
By the end of 60 epochs, the values for MAE and NRMSE are $0.088$ and $0.059$, respectively, suggesting that some pixel-by-pixel differences between the images still remain.

The MS-SSIM and SSIM metrics show steady improvement over time, indicating excellent visual similarity between the generated and target high-resolution images, as we have already qualitatively seen in Fig.~\ref{fig:epoch_60}.
At the end of 60 epochs, their values are $0.962$ and $0.841$, which are notably high.

The PSNR is also increasing, reaching a value of $24.605$ after 60 epochs, indicating that images are of good quality in terms of their signal strength.

Alternatively, the loss function can be changed from mean squared error (MSE) to mean absolute error (MAE) to reassess the model performance.
Since MSE penalizes larger error values more heavily due to its squared nature, MAE, which treats all error values equally might provide a balanced training. 
Listing~\ref{cd:modify_loss_function} shows the updated loss function as a modification to Listing~\ref{cd:loss_function}. 

\begin{lstlisting}[
    label=cd:modify_loss_function,
    caption=Defining the loss function (by modifying Listing~\ref{cd:loss_function})
]
criterion = torch.nn.L1Loss()
\end{lstlisting}
After 60 epochs with the modified loss, the MAE and NRMSE metrics show a significant improvement, with values of $0.064$ and $0.047$ respectively.
The MS-SSIM, SSIM, and PSNR metrics, which assess the visual quality of images, have largely remained the same with only minor improvements, with the values $0.962$ $0.849$ and $26.475$, respectively.

To improve both the visual and quantitative benchmarks, experimenting with different learning schedules, including variable learning rates during training~\cite{Maggiora23_ConditionalVariational}, may enhance model performance.
Alternatively, techniques such as SR3~\cite{Saharia21_ImageSuperResolution} and Latent Diffusion Models~\cite{Rombach22_HighResolutionImage} can also be employed.
Each of these strategies excels in optimizing different metrics, making it crucial to combine the quantitative metrics with careful visual inspection to achieve the best results for the task at hand.
It is important to note that image super-resolution is an ill-posed inverse problem~\cite{yang2019deep}. 
This means that, since the low-resolution image lacks significant details, there could be many possible high-resolution images corresponding to a single low-resolution image. 
Consequently, there are inherent theoretical limitations that prevent any deep learning model from perfectly reconstructing the target image in every detail.

\section{Conclusion}\label{sec:conclusion}

In this tutorial, we have explored how to implement a denoising diffusion probabilistic model (DDPM) from scratch, covering both the mathematical foundations and the corresponding implementation in Python.
We applied this model to the task of generating super-resolution images.
Using the BioSR dataset's microtubule structure, we implemented a dataset class to define the augmentations and transformations, created a diffusion class to handle the forward and reverse diffusion processes, and finally, built a training loop to generate high-resolution images from the low-resolution images test images.
Additionally, we discussed alternative implementations and strategies to enhance the model's performance. 

With minimal adjustments, this tutorial can be adapted to a wide range of tasks beyond super-resolution. For instance, the same principles can be applied to generate entirely new images from noise, a process known as unconditional image generation.
To do this, simply omit the low-resolution images from the reverse diffusion process, which was previously used as a condition along with the noise (Section~\ref{sec:implement_reverse_diffusion}), and change the number of input channels in the attention U-Net from $2$ to $1$ (Listing~\ref{cd:Attention_U-Net}).
By excluding them, the reverse diffusion process will start with pure noise and generate unseen high-resolution images that are unrelated to any low-resolution input, while maintaining the characteristics of the dataset.
Although this approach is not particularly suited for image-to-image translation task like super-resolution, it is valuable for generating new unseen data, especially in cases where the dataset is limited.
For instance, it can be used to augment training data, create synthetic datasets, or evaluate the generative capabilities of the model in producing novel and high-quality images.

This tutorial can also be adapted for other conditional image generation tasks such as text-to-image generation, and class-conditioned imaged generation.
For text-to-image generation, a text prompt can be provided in its encoded format directly to the model, serving as a condition.
To achieve this, text prompts need to be generated for each image in the dataset.
For instance, in a dataset like MNIST, a prompt would look like ``Get an image of an eight''.
This prompt is then passed through a text tokenizer (a dictionary that assigns numbers to words in a text prompt), and subsequently processed by a transformer~\cite{Vaswani17_AttentionAll} to generate a text encoding.
The text encoding can be incorporated at different levels within the attention U-Net, similar to the time step (Section~\ref{sec:attention_unet}).
A detailed example of this approach can be found in reference~\cite{Volpe25_DeepLearningCrash}.

The same principle applies to class-conditioned generation, such as generating images of specific biological structure in the BioSR dataset or generating an image from the MNIST dataset based on a class label.
The class label is transformed into an encoding and then incorporated at various levels within in attention U-Net.
For both text-to-image and class-conditioned models, more nuanced approaches such as classifier-free guidance~\cite{ho2021classifierfree} can be employed, where the conditions are randomly dropped occasionally to preserve both the conditional and unconditional generational capabilities of the model.
For image-to-image translation tasks such as virtual staining of microscopy images, or transforming microscopy modalities from one domain to another, or in-painting, the tutorial operates in much the same way, with the primary change being the dataset used.

\appendix
\section*{Appendix}

\section{Derivation of Fast Forward Process}\label{appendix:fast_forward}

In this section, we will derive the fast forward equation (Eq.~\ref{eq:fastforward}) starting from the forward diffusion process (Eq.~\ref{eq:forward_diffusion}).
As we have already seen, the fast forward equation has allowed us to directly jump from the initial image to the image at time $t$, without having to generate all the intermediate noisy images.
From Eq.~\ref{eq:forward_diffusion}, we have
\begin{equation}
	{\bf x}_t = \sqrt{1-\beta_t} \, {\bf x}_{t-1} + \sqrt{\beta_t} \, \boldsymbol{\epsilon}_{t-1}
\end{equation}
Rewriting it in terms of $\alpha_t = 1 - \beta_t$, we get
\begin{equation}
	{\bf x}_t = \sqrt{\alpha_t} \, {\bf x}_{t-1} + \sqrt{1-\alpha_t} \, \boldsymbol{\epsilon}_{t-1}
\end{equation}
By expanding {\bf x}$_{t-1}$ in terms of {\bf x}$_{t-2}$, we have
\begin{eqnarray}\label{eq:fd_expand}
    {\bf x}_t 
    & = & 
    \sqrt{\alpha_t} \left(\sqrt{\alpha_{t-1}} \, {\bf x}_{t-2} + \sqrt{1-\alpha_{t-1}} \, \boldsymbol{\epsilon}_{t-2}\right) + \sqrt{1-\alpha_t} \, \boldsymbol{\epsilon}_{t-1} \nonumber \\
    & = & 
    \sqrt{\alpha_t \alpha_{t-1}} \, {\bf x}_{t-2} + \sqrt{\alpha_t (1-\alpha_{t-1})} \, \boldsymbol{\epsilon}_{t-2} + \sqrt{1-\alpha_t} \, \boldsymbol{\epsilon}_{t-1}
\end{eqnarray}
In the above equation, the last two terms can be viewed as random variables sampled from two distinct Gaussian distributions, one with mean $0$ and variance $\alpha_t (1-\alpha_{t-1})$ (for the penultimate term) and the other with mean $0$ and variance $1-\alpha_t$ (for the final term), respectively.
Notice also the use of reparametrization trick here.
In addition, the summation stability of Gaussian distributions states that the sum of two independent Gaussian random variables is also a Gaussian random variable that has been sampled from a Gaussian distribution whose mean and variance are equal to the sum of the means and variances, respectively, of the two random variables.

Combining these two observations, we can merge the last two terms of Eq.~\ref{eq:fd_expand} into a single term, which represents a random variable sampled from a Gaussian distribution with a mean $0$ and variance $\alpha_t (1-\alpha_{t-1}) + 1-\alpha_t = 1 - \alpha_{t-1} \alpha_t$. Consequently, we obtain
\begin{equation}
	{\bf x}_t = \sqrt{\alpha_t \alpha_{t-1}} \, {\bf x}_{t-2} + \sqrt{1 - \alpha_{t-1} \alpha_t} \, \boldsymbol{\epsilon_t}
\end{equation}
where $\boldsymbol{\epsilon_t}$ is still a random variable sampled from a standard Normal distribution with mean $0$ and variance $1$. 

If we repeat the above process recursively, we get
\begin{eqnarray}
    {\bf x}_t 
    & = & 
    \sqrt{\alpha_t \alpha_{t-1} \dots \alpha_1} \, {\bf x}_0 + \sqrt{1 - \alpha_1 \alpha_2 \dots \alpha_t} \, \boldsymbol{\epsilon_t} \nonumber \\
    {\bf x}_t 
    & = & 
    \sqrt{\bar\alpha_t} \, {\bf x}_0 + \sqrt{1 - \bar\alpha_t} \, \boldsymbol{\epsilon_t}
\end{eqnarray}
where $\bar\alpha_t = \prod_{s=1}^t \alpha_s$ and $\alpha_t = 1 - \beta_t$.
This is how we arrive at the fast forward equation in Eq.~\ref{eq:fastforward}.

psection{Simplification of Posterior Mean}\label{appendix:posterior_mean}

Here, we will derive the simplified form of posterior mean as shown in Eq.~\ref{eq:simplified_posterior_mean}. 
From the unsimplified posterior mean (Eq.~\ref{eq:posterior_mean}) and the fast forward equation (Eq.~\ref{eq:fastforward}), we have
\begin{equation}
    \tilde{\boldsymbol{\mu}} ({\bf x}_t, {\bf x}_0)
    =
    \frac{\sqrt{\bar{\alpha}_{t - 1}} \ \beta_t}{1 - \bar{\alpha}_t} \ {\bf x}_0
    +
    \frac{\sqrt{\alpha_t} (1 - \bar{\alpha}_{t-1})}{1 - \bar{\alpha}_t} \ {\bf x}_t,
\end{equation}
and
\begin{equation}
    {\bf x}_0 
    = 
    \frac{{\bf x}_{t} - \sqrt{1-\bar\alpha_t} \, \boldsymbol{\epsilon}_t}{\sqrt{\bar\alpha_t}}
\end{equation}
By plugging the second equation into the first and collecting the coefficients of ${\bf x}_t$, we get
\begin{equation}
    \tilde{\boldsymbol{\mu}}
    =
    \left(\,
        \frac{\sqrt{\bar{\alpha}_{t - 1}} \, \beta_{t} }{(1 - \bar{\alpha}_t)} \,
        \frac{1}{\sqrt{\bar\alpha_t}} 
        +
        \frac{\sqrt{\alpha_t} (1 - \bar{\alpha}_{t-1})}{1 - \bar{\alpha}_t} \,   
    \right) \,
    {\bf x}_t \,
    -
    \frac{\beta_{t} \, \sqrt{\bar{\alpha}_{t - 1}} \, \sqrt{1-\bar\alpha_t}}{(1 - \bar{\alpha}_t) \, \sqrt{\bar\alpha_t}} \,
    {\boldsymbol{\epsilon}}_t
\end{equation}
Now, by substituting $\beta_{t} = 1 - \alpha_{t}$ and $\bar{\alpha}_{t} = \alpha_{t} \, \bar{\alpha}_{t-1}$, we get
\begin{equation}
    \tilde{\boldsymbol{\mu}}
    =
    \left(
        \frac{(1 - \alpha_{t})}{1 - \bar{\alpha}_t} \,
        \frac{1}{\sqrt{\alpha_t}} 
        +
        \frac{\sqrt{\alpha_t} (1 - \bar{\alpha}_{t-1})}{1 - \bar{\alpha}_t}    
    \right) \,
    {\bf x}_t \,
    -
    \frac{(1 - \alpha_{t})}{\sqrt{\alpha_{t}} \, \sqrt{1 - \bar{\alpha_{t}}}} \,
    {\boldsymbol{\epsilon}}_t
\end{equation}
By simplifying the first term and combining both the terms, we finally get
\begin{equation}
    \tilde{\boldsymbol{\mu}}
    =
    \left(
        \frac{(1 - \alpha_{t}) + \alpha_t (1 - \bar{\alpha}_{t-1})}{\sqrt{\alpha_t} \, (1 - \bar{\alpha}_t) } 
    \right) \,
    {\bf x}_t \,
    -
    \frac{(1 - \alpha_{t})}{\sqrt{\alpha_{t}} \, \sqrt{1 - \bar{\alpha_{t}}}} \,
    {\boldsymbol{\epsilon}}_t
\end{equation}
and
\begin{equation}
    \tilde{\boldsymbol{\mu}}
    =
    \left(
        \frac{1}{\sqrt{\alpha_t}} 
    \right) \,
    {\bf x}_t \,
    -
    \frac{(1 - \alpha_{t})}{\sqrt{\alpha_{t}} \, \sqrt{1 - \bar{\alpha_{t}}}} \,
    {\boldsymbol{\epsilon}}_t
\end{equation}
In this way, we can obtain the simplified form of the posterior mean as shown in Eq.~\ref{eq:simplified_posterior_mean}:
\begin{equation}
    \boxed{
    \tilde{\boldsymbol{\mu}}
    =
    \frac{1}{\sqrt{\alpha_{t}}}
    \left(
        {\bf x}_t \,
        -
        \frac{1 - \alpha_{t}}{\sqrt{1 - \bar{\alpha_{t}}}} \,
        {\boldsymbol{\epsilon}}_t
    \right)
    }
\end{equation}

\section{Alternative Implementation of Reverse Process}\label{appendix:alt_reverse_diffusion}

In the previous implementation of the reverse diffusion process in Sections~\ref{sec:reverse_diffusion} and \ref{sec:implement_reverse_diffusion}, we have used the recursive equation (Eq.~\ref{eq:reverse_diffusion}) that was based on the simplified posterior mean (Eq.~\ref{eq:simplified_posterior_mean}). In this alternative implementation, we will use the original form the posterior mean, namely,
\begin{equation}\label{eq:posterior_mean_coeff}
    \boldsymbol{\mu}_{\theta} ({\bf x}_t, {\bf x}_0)
    =
    \underbrace{\frac{\sqrt{\bar{\alpha}_{t - 1}} \ \beta_t}{1 - \bar{\alpha}_t}}_{\mbox{coefficient $1$}} \, {\bf x}_0 \,
    + \,
    \underbrace{\frac{\sqrt{\alpha_t} (1 - \bar{\alpha}_{t-1})}{1 - \bar{\alpha}_t}}_{\mbox{coefficient $2$}} \, {\bf x}_t,
\end{equation}
where
\begin{equation}\label{eq:clean_image_coeff}
    {\bf x}_0 
    = 
    \underbrace{\sqrt{\frac{1}{\bar\alpha_t}}}_{\mbox{coefficient $1$}} {\bf x}_t
    - \,
    \underbrace{\sqrt{\frac{1}{\bar\alpha_t} - 1}}_{\mbox{coefficient $2$}} \, 
    \boldsymbol{\epsilon}_{\theta}({\bf x}_t, t)
\end{equation}\\
is the clean image reconstructed from the forward diffusion equation (Eq.~\ref{eq:forward_diffusion}). Notice that the noise $\boldsymbol{\epsilon}_{\theta}({\bf x}_t, t)$ is now the noise predicted by the neural network.
In every time step, we will use this predicted noise to reconstruct the clean image ${\bf x}_0$, and use it to calculate the posterior mean.
To make the implementation of the code easy, we label the different parts of both the equations with coefficient numbers.

For the posterior variance, we can either use $\boldsymbol{\Sigma}_\theta = \sigma^2_t = \beta_t$ or $\sigma^2_t = \tilde\beta_t$, where $\tilde\beta_t = \frac{1 - \bar\alpha_{t-1}}{1-\bar\alpha_t} \, \beta_t$ (Section~\ref{sec:reverse_diffusion}). 
In this alternative implementation, we will use the second choice (we have used the first choice in the previous implementation in Sections~\ref{sec:reverse_diffusion} and \ref{sec:implement_reverse_diffusion}). 
Since we now have both the posterior mean and variance, we can use the reparametrization trick to arrive at the recursive equation ${\bf x}_{t-1} = \boldsymbol{\mu}_{\theta} ({\bf x}_t, {\bf x}_0) + \sqrt{\tilde\beta_t} \, {\bf z}$, where ${\bf z}$ is a pure noise image sampled from a standard Normal distribution (${\bf z} \sim \mathcal{N}(0, 1)$).

We are now ready to implement this into the \lstinline{Diffusion} class, as shown in Listing~\ref{cd:alternative_reverse_diffusion}, by modifying Listing~\ref{cd:reverse_diffusion}.
\begin{lstlisting}[
    label=cd:alternative_reverse_diffusion,
    caption=Alternative implementation of the reverse diffusion process (by modifying Listing~\ref{cd:reverse_diffusion})
]
€€€class Diffusion:
    """Denoising diffusion probabilstic model (DDPM)."""
    
    def __init__(self, noise_steps=1000, beta_start=1e-4, beta_end=0.02,
        img_size=128, device=device):€€€

        €€€. . . . . .€€€
        
        self.alpha_bar_prev_step = torch.cat( (@\circled{1}@)
            (torch.tensor([1.0]).to(device), self.alpha_bar[:-1]))

        # posterior mean coefficients (Eq. 26)
        self.posterior_mean_coeff1 = (
            (torch.sqrt(self.alpha_bar_prev_step) * self.beta) / (1 - self.alpha_bar)
        )[:, None, None, None] (@\circled{2}@)
        self.posterior_mean_coeff2 = (
            (torch.sqrt(self.alpha) * (1 - self.alpha_bar_prev_step))
            / (1 - self.alpha_bar)
        )[:, None, None, None] (@\circled{3}@)

        # forward diffusion coefficients (Eq. 27)
        self.x0_coeff1 = torch.sqrt(1 / self.alpha_bar)[:, None, None, None] (@\circled{4}@)
        self.x0_coeff2 = torch.sqrt(1 / self.alpha_bar - 1)[:, None, None, None] (@\circled{5}@)

        # posterior variance
        variance = self.beta * (1 - self.alpha_bar_prev_step) / (1 - self.alpha_bar) (@\circled{6}@)
        self.posterior_log_variance_clipped = torch.log(
            torch.max(variance, torch.ones_like(variance) * 1e-20)
        )[:, None, None, None] (@\circled{7}@)



    €€€. . . . . .€€€

    def reconstruct_clean_image(self, x_t, t, noise): (@\circled{8}@)
        """Returns the reconstructed clean image"""
        return self.x0_coeff1[t] * x_t - self.x0_coeff2[t] * noise

    def posterior_parameters(self, x0, x_t, t): (@\circled{9}@)
        """Returns posterior mean and variance (in log form)"""
        posterior_mean = (
            self.posterior_mean_coeff1[t] * x0 
            + self.posterior_mean_coeff2[t] * x_t
        )
        posterior_log_variance_clipped = self.posterior_log_variance_clipped[t]
        return posterior_mean, posterior_log_variance_clipped

    €€€def reverse_diffusion(self, model, n_images, n_channels,
                          position_encoding_dim, position_encoding_function,
                          save_time_steps=None, input_image=None):
        """Reverse diffusion process"""

        with torch.no_grad():
            x = torch.randn((n_images, n_channels, self.img_size, self.img_size))
            x = x.to(self.device)
            
            denoised_images = []
            for i in tqdm(reversed(range(0, self.noise_steps)),
                          desc="U-Net inference", total=self.noise_steps):

                t = (torch.ones(n_images) * i).long()
                t_pos_enc = position_encoding_function(
                    t.unsqueeze(1), position_encoding_dim
                ).to(self.device)

                if input_image is None:
                    predicted_noise = model(x=x, t=t_pos_enc)
                else:
                    predicted_noise = model(
                        torch.cat((input_image.to(self.device), x), dim=1),
                        t_pos_enc,
                    )€€€
 
                &&&alpha = self.alpha[t][:, None, None, None]&&&
                &&&alpha_bar = self.alpha_bar[t][:, None, None, None]&&&    

                x0_reconstructed = self.reconstruct_clean_image( (@\circled{10}@)
                    x, t, noise=predicted_noise
                )
                x0_reconstructed = x0_reconstructed.clamp(-1, 1)

                posterior_mean, posterior_log_variance = self.posterior_parameters( (@\circled{11}@)
                    x0=x0_reconstructed, x_t=x, t=t
                )
                
                €€€if i > 0:
                    noise = torch.randn_like(x)
                else:
                    noise = torch.zeros_like(x)€€€
                
                &&&x = (1 / torch.sqrt(alpha) * (x - ((1 - alpha)&&& 
                    &&&/ torch.sqrt(1 - alpha_bar)) * predicted_noise)&&& 
                    &&&+ torch.sqrt(1-alpha) * noise)&&&

                x = posterior_mean + noise * torch.exp(0.5 * posterior_log_variance) (@\circled{12}@)

                €€€if i in save_time_steps:
                    denoised_images.append(x)

            denoised_images = torch.stack(denoised_images)
            denoised_images = denoised_images.swapaxes(0, 1)
            return denoised_images€€€
\end{lstlisting}
The diffusion class is now initialized with several new parameters.
Starting with the \lstinline{alpha_bar_prev_step} parameter \circled{1} representing $\bar\alpha_{t-1}$, it initializes the  \lstinline{posterior_mean_coeff1} and \lstinline{posterior_mean_coeff2}  parameters \circled{2}\circled{3}, which represent the posterior mean coefficients  (Eq.~\ref{eq:posterior_mean_coeff}).
It then initializes clean image coefficients (Eq.~\ref{eq:clean_image_coeff}) through the \lstinline{x0_coeff1} and \lstinline{x0_coeff2} parameters \circled{4}\circled{5}.

The posterior variance $\tilde\beta$ is defined through the \lstinline{variance} parameter \circled{6}.
To ensure numerical stability for lower variance values, a logarithmic version of the variance is calculated through the  \lstinline{posterior_log_variance_clipped} parameter \circled{7}. 
Before applying the logarithm, the variance is clipped between the actual value and a minimum value to avoid undefined logarithm operations.

The \lstinline{reconstruct_clean_image()} and \lstinline{posterior_parameters()} methods \circled{8}\circled{9} implement Eq.~\ref{eq:clean_image_coeff} and Eq.~\ref{eq:posterior_mean_coeff} using the previously initialized parameters.

Finally, the reverse diffusion is modified to accommodate the alternative implementation.
The implementation is very similar to the earlier implementation in Listing~\ref{cd:reverse_diffusion} with one major distinction. 
In every step, instead of directly calculating the (slightly) denoised image ${\bf x}_t$, it first reconstructs the clean image ${\bf x}_0$ \circled{10} using predicted noise, calculates the posterior mean and variance \circled{11}, and with the help of reparametrization trick it arrives at the denoised image ${\bf x}_t$ \circled{12}. 
This elaborate approach is observed to produce better results with color images where multiple channels are involved.
This modified implementation of \lstinline{Diffusion} class works seamlessly with the rest of this tutorial, requiring no additional changes. 

\section*{Acknowledgments}
The authors acknowledge the support from the Horizon Europe ERC Consolidator Grant MAPEI (Grant No. 101001267) and the Knut and Alice Wallenberg Foundation (Grant No. 2019.0079).

\bibliography{bibliography}

\end{document}